\documentclass[review]{elsarticle}

\usepackage{lineno}
\modulolinenumbers[5]
\usepackage{graphicx}
\usepackage{amssymb}
\usepackage{booktabs}
\usepackage{multirow}
\usepackage{subfigure}
\usepackage{stfloats}
\usepackage{tabularx}
\usepackage{wrapfig}
\usepackage[colorlinks=true,
linkcolor=black]{hyperref}

\begin{document}

\begin{frontmatter}

\title{Modeling the Social Influence of COVID-19 via Personalized Propagation with Deep Learning}

\author[mymainaddress,mysecondaryaddress]{Yufei Liu}
\ead{liuyufei@nuaa.edu.cn}

\author[mymainaddress]{Jie Cao\corref{mycorrespondingauthor}}
\cortext[mycorrespondingauthor]{Corresponding author}
\ead{caojie690929@njust.edu.cn}

\author[mysecondaryaddress]{Dechang Pi}
\ead{pinuaa@nuaa.edu.cn}

\address[mymainaddress]{College of Information Engineering, Nanjing University of Finance and Economics, Nanjing 210023, China}
\address[mysecondaryaddress]{College of Computer Science and Technology, Nanjing University of Aeronautics and Astronautics (NUAA), Nanjing 211106, China}

\begin{abstract}
Social influence prediction has permeated many domains, including marketing, behavior prediction, recommendation systems, and more. However, traditional methods of predicting social influence not only require domain expertise, they also rely on extracting user features, which can be very tedious. Additionally, graph convolutional networks (GCNs), which deals with graph data in non-Euclidean space, are not directly applicable to Euclidean space. To overcome these problems, we extended DeepInf such that it can predict the social influence of COVID-19 via the transition probability of the page rank domain. Furthermore, our implementation gives rise to deep learning-based personalized propagation algorithm, called DeepPP. The resulting algorithm combines the personalized propagation of a neural prediction model with the approximate personalized propagation of a neural prediction model from page rank analysis. Four social networks from different domains as well as two COVID-19 datasets were used to demonstrate the efficiency and effectiveness of the proposed algorithm. Compared to other baseline methods, DeepPP provides more accurate social influence predictions. Further, experiments demonstrate that DeepPP can be applied to real-world prediction data for COVID-19.
\end{abstract}

\end{frontmatter}
%\linenumbers
%\newpage
\section{Introduction}
In the current era of big data, social networks are providing a plethora of information on user interactions \cite{LU2016143,WANG2017154}. Twitter, for example, had 206 million daily active users in the second quarter of 2021 \cite{DAUD2020102716}. However, what we are seeing with the isolation imposed by COVID-19 is social networks playing an even more important role in the everyday interactions between people - particularly when it comes to social influence. The term `social influence' is usually understood as the process whereby a user's emotions, opinions, or behaviors are shaped by their environment, i.e., the process by which people alter their behavior under the influence of others \cite{SINGH2021151}. With the globalization of online social networks, social influence analysis has spread to many domains, including marketing \cite{DU2022386}, behavior prediction \cite{DAUD2020102716}, recommendation systems \cite{liu2018social,Gao_CIKM16}, influence maximization \cite{RAJ202236}, public opinion guiding \cite{YANG2021107640}, communities \cite{ijcai2020-693}, and graph anomaly detection \cite{ma2021comprehensive}. We cannot deny that social influence has become ubiquitous and complex in shaping our social decisions. Therefore, there is much interest in developing methods to understand, describe, and identify the mechanisms and evolutions behind the social influence.

The process of predicting influence on a user involves sampling local neighbors, building a local network from the samples, and then learning the potential predictive signals from that network. Matsubara et al. \cite{matsubara2012rise} designed a dynamic model of social influence based on a differential equation drawn from classic susceptibility theory. Li et al. \cite{li2017deepcas} proposed an end-to-end predictor that infers the cascading size of features by combining RNNs with representation learning. With both these methods, the aim is to predict the statistical patterns of social influence over time, including cascading size and global patterns. Qiu et al. \cite{qiu2018deepinf} proposed the famous social influence prediction method, DeepInf, designed for general social network settings. The aim is to predict a user's behavior state by considering the neighbor's behavior state and characteristics. However, due to the sparsity, the neighbor information of the COVID-19 network is very limited, so they are invalid for the network-based COVID-19 prediction.

\begin{figure}[htb]
\centering
\subfigure[]{
\label{Fig:sub:1}
\includegraphics[width=0.4\textwidth]{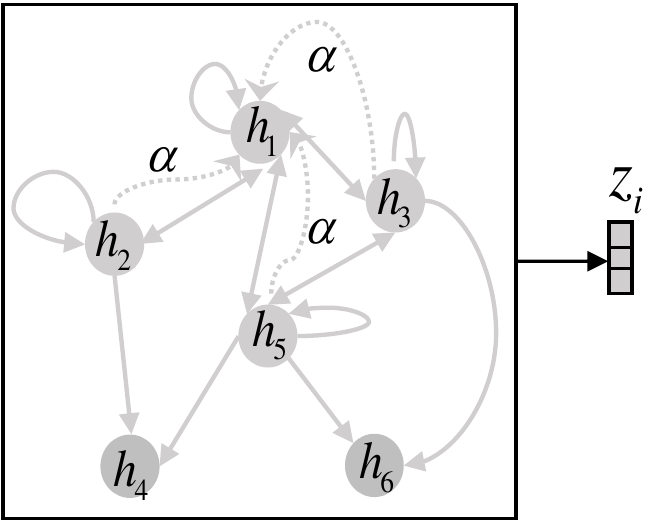}}
\subfigure[]{
\label{Fig:sub:2}
\includegraphics[width=0.4\textwidth]{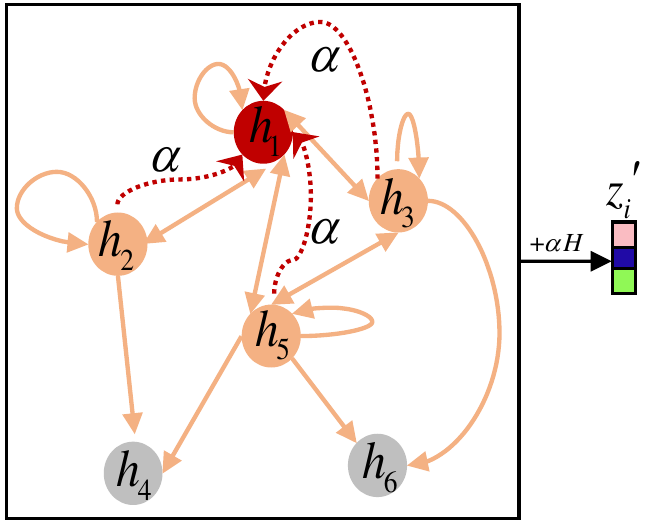}}
\caption{The neural network first predicts the influence of each node according to its own characteristics. Then, the personalized PageRank algorithm is used to propagate influence adaptively. Panel (a) shows the personalized propagation. Panel (b) shows the improved personalized propagation.}
\label{fig:1}
\end{figure}

In this paper, to overcome these issues, we extended DeepInf such that our proposed model utilizes personalized propagation to predict social influence at the user level. We built on DeepInf \cite{qiu2018deepinf} by integrating the transition probability $\alpha$ of the page rank domain with a model based on a graph convolutional network (GCN), as shown in Figure~\ref{fig:1}. Here, $\alpha$ adjusts the size of neighborhood influence, $\mathbf{H}$ is the prediction function, and $\alpha \mathbf{H}$ offers greater flexibility. More specifically, we extended DeepInf by leveraging the two algorithms, PPNP \cite{klicpera2018personalized} (personalized propagation of neural predictions) and APPNP \cite{klicpera2018personalized} (approximation to PPNP) extended of a GCN, and transplanting them from page ranking into social influence analysis. Our implementation gives rise to deep learning-based personalized propagation algorithm, called DeepPP. The algorithm predicts how $v$ will behave in the future based on its current behavior by analyzing the state of $n$ neighbors around a user $v$. Nodes (users) in the system can have one of two states: active or inactive. But, to forecast the state $v$ at the end of the interval, we need to know $v$'s current state.

To verify the efficiency and effectiveness of our algorithm, we used four social networks in various domains: Open Academic Graph (OAG), Digg, Twitter, and Weibo, as well as two COVID-19 datasets (Hubei and Holland). We compared DeepPP with: a conventional GNN* model \cite{4700287}; PPNP and APPNP \cite{klicpera2018personalized}; and the state-of-the-art models DeepInf \cite{qiu2018deepinf} and DeepEmLAN \cite{ZHAO2021382}. The results of these experiments showed that DeepPP is more accurate than the GNN* method and, overall, is superior to all the other algorithms in terms of social influence prediction accuracy.

The following list summarizes our main contributions.
\begin{enumerate}
  \item Inspired by DeepInf, we integrated the transition probability $\alpha$ of the page rank domain into a graph convolutional network (GCN) model, thereby extending DeepInf.
  \item The resulting DeepPP algorithm repurposes PPNP and APPNP from page ranking to social influence analysis without any additional time complexity.
  \item Comprehensive evaluations of model performance show DeepPP to be more accurate than the baseline methods.
\end{enumerate}

The remainder of this paper proceeds as follows. In Section 2, we provide a brief review of the status of research on social influence prediction. Section 3 introduces the overview of DeepPP and outlines the personalized propagation process based on deep learning for modeling the social influence of COVID-19. Section 4 describes the experiments used to validate DeepPP. Concluding remarks and future research directions are highlighted in Section 5.

\section{Related work}
Our literature review covers the following topics: traditional social influence analysis, deep learning-based social influence prediction, and graph representation learning. The results show that an accurate model for predicting social influence may not have been established.

\subsection{Methods of Traditional Social Influence Analysis}
Social influence analysis is traditionally based on the study of interpersonal communications, where user features are typically extracted by hand, which can be tedious. Further, the analyses are based on domain expertise in sociology or cognitive science and, therefore, extending the results is difficult \cite{qiu2018deepinf}. A study by Li et al. \cite{li2018social} distinguished two types of social influence analysis models: (1) macro, which assumes all users have equal power to influence; and (2) micro, which explores individual levels of user influence. Of the micro-level influence models, independent cascade and linear threshold are the two most common. Nevertheless, both types of models assume that users will not change their state, i.e., the probability of influencing others or being influenced \cite{Koskinen} using Bayes Theorem \cite{WU20151487,wu2011attribute}. As such, they are not particularly reflective of real social networks.

\subsection{Methods of social influence prediction based on deep learning}
It is well known that deep learning has benefitted many fields, and social influence analysis is no exception. However, its application to this field is relatively recent. To date, both micro- and macro-level methods of analysis have been developed. Micro-level models focus on user interactions. The key assumption is that user interactions, e.g., ratings, comments, retweets, affect the behavior of others. DeepInf \cite{qiu2018deepinf}, one of the current state-of-the-art models, is an end-to-end model for predicting user-level influence. It integrates a network embedding \cite{liao2018attributed} with a graph attention network (GAT) \cite{wang2019heterogeneous} and a GCN \cite{wu2019simplifying}. Both GCNs and GATs are better suited to non-Euclidean spatial data at aggregating features of neighbor vertices onto central vertices than conventional machine learning methods, such as Gradient Boosted Decision Tree (GBDT) and \textit{K}-Nearest Neighbor (KNN) \cite{zhang2018salient}. In fact, experiments with DeepInf show the best predictive performance with multi-headed GATs, even when compared to a GCN \cite{wu2019simplifying}. By using dual GNNs instead of a single GNN in recommender systems, Wu et al. \cite{wu2019dual} designed a deep latent representation of multifaceted social impact.

The macro-level solutions focus on the patterns of global social influence. For example, DeepCas  \cite{li2017deepcas}, a method of analyzing macro-social impact models using RNNs, involves a pattern of cascading that includes all aspects of the cascading and their associations with the final cascading size \cite{liao2018attributed}. In this method, as a two-dimensional colored diagram, an end-to-end predictor visualizes all cascades of influence information. Using DeepCas as inspiration, Cao et al. \cite{cao2017deephawkes} developed a method which represents information cascades using an explainable model of the generation Hawkes process \cite{menon2018proper}. Their model showed better performance than traditional generation and feature-based models. DeepHawkes \cite{cao2017deephawkes}, as well as DeepInf, is data-driven, which means it can learn from previous cascades and can take advantage of historical information. A data-driven cascading method called LSTMICs, that uses long short-term memory (LSTM), was further developed and extended by Gou et al. \cite{gou2018learning} to learn sequence features from cascading features and RNN functions. They used a Weibo and a Twitter dataset to predict outbreaks more accurately than previous methods. With Cas2vec \cite{kefato2018cas2vec}, it is possible to accurately predict virus cascades without manually extracting the relevant information from the framework, which could be costly to obtain. It is based on the time interval between two events rather than the information contained in those events that extracts are made.

 Social influence analysis models are closely related to graph representation learning, and many studies have been conducted on graph representation learning as part of graph mining, e.g., DeepWalk \cite{perozzi2014deepwalk}, Line \cite{wang2017community}, Node2vec \cite{grover2016node2vec}, Metapath2vec \cite{dong2017metapath2vec}, NetMF \cite{liu2019general}, Graph Kernel \cite{siglidis2020grakel}, and the most advanced method PSCN \cite{niepert2016learning}. More recently, researchers have been exploring the notion of graph representation learning with semi-supervised information. Some examples include GCN \cite{wu2019simplifying}, GraphSage \cite{Hamilton_NIPS}, and GAT \cite{wang2019heterogeneous} as the most advanced model.

\section{Proposed model}
\begin{figure*}[htbp]
\begin{center}
\includegraphics*[width=1\textwidth]{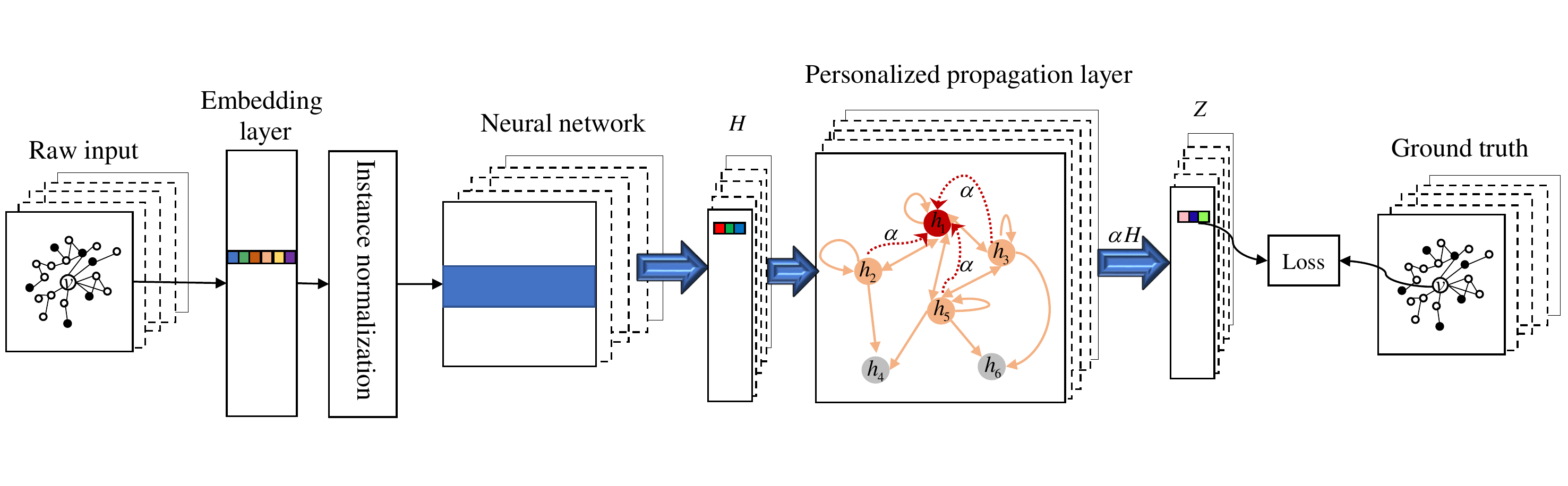}
\caption{Overview of the DeepPP model. The neighbors of node $v$ are sampled with the goal being to predict the state of node $v$ after a number of iterations.}
\label{fig:2}
\end{center}
\end{figure*}

Analyzing social influence with deep learning techniques is a problem related to graph neural networks. Because graph neural networks in recommendation systems have other types of applications, such as collaborative filtering and predictive page ranking, drawing inspiration from other fields to improve our existing research is a highly useful exercise. Figure \ref{fig:2} shows an overview of personalized propagation based on deep learning for modeling the social influence associated with COVID-19.

\subsection{Influence propagating}
We extended DeepInf by leveraging two algorithms: personalized propagation of neural predictions (PPNP) and an approximation to PPNP (APPNP). Klicpera et al. \cite{klicpera2018personalized} proposed them by extending a GCN, where each node has the same effect on its neighbors. In other words, the independent cascade model recalled in Section 2 has the same influence scores and probability of being propagated to its neighbors for each edge. With the transfer probability $\alpha$, information transmission is adjusted so that each node has a different influence on its neighbors. In experiments, PPNP and APPNP algorithms perform better than the GNN algorithm when the transfer probability $\alpha$ is set to between 0.05 and 0.2 \cite{klicpera2018personalized}. This is also compared to the GCN algorithm, which has an average effect on all neighbors. The transport vector $\mathbf{i}_x$ preserves the local neighborhood of the node \cite{ksantini2007weighted}. Essentially, the $I(x,y)$ score (i.e. root node $x$ influence on node $y$) is equivalent to the $y^{(th)}$ element of a personalized page rank $\pi_{pr}(\mathbf{i}_x)$. A recursive equation with a transfer vector is expressed with a transfer probability $\alpha\in(0,1)$ as:

\begin{equation}
\pi _{pr}\left( {{\mathbf{i}_x}} \right) = \alpha {\left( {{\mathbf{I}_n} - \left( {1 - \alpha } \right)\hat \mathbf{A}} \right)^{ - 1}}{\mathbf{i}_x}
\end{equation}

The influence score in Equation (1) can be used to generate the prediction objective function.
\begin{equation}
Z = softmax\left( {\alpha {{\left( {{\mathbf{I}_n} - \left( {1 - \alpha } \right)\hat \mathbf{A}} \right)}^{ - 1}}\mathbf{H}} \right), {\mathbf{H}_{i,i}} = {f_\theta }\left( {{\mathbf{X}_{i,s}}} \right)
\end{equation}
where $\mathbf{X}_i$ is the eigenmatrix. $f_\theta$ is a neural network that predicts $\mathbf{H}\in\mathbb{R}^{n\times c}$.
Equations (1) and (2) are the PPNP model. However, calculating the time complexity requires a dense matrix $\mathbb{R}^{n\times c}$ with the time complexity $O(n^2)$. APPNP was developed as a means of addressing this shortcoming. More specifically, by using the power method, which has a linear computational complexity, the eigenvalues for diagonalizable matrices are calculated as follows:
\begin{equation}
Z^{\left( 0 \right)} = \mathbf{H} = {f_\theta }\left( \mathbf{X} \right)
\end{equation}
\begin{equation}
{Z^{\left( {k + 1} \right)}} = \left( {1 - \alpha } \right)\hat \mathbf{A}{Z^{\left( k \right)}}{\rm{ + }}\alpha \mathbf{H}
\end{equation}
\begin{equation}
{Z^{\left( K \right)}} = softmax\left( {\left( {1 - \alpha } \right)\hat \mathbf{A}{Z^{\left( {K - 1} \right)}} + \alpha \mathbf{H}} \right)
\end{equation}
where $Z^{(k)}$ is a transport set that can effectively provide approximate predictions.

\subsection{Network encoding with a GCN}

Our model is derived from a GCN and a graph attention network model as follows.

GCN, that analyses graph-structured data, is semi-supervised learning \cite{wu2019simplifying}. For eigen decomposition, the GCN uses a Laplace graph of the Fourier domain. Specifically, the rule in Equation (6) for a GCN is layer-by-layer propagation, which consists of multiple GCN layers stacked on top of each other.

\begin{equation}
f\left( {{\mathbf{H}^{\left( l \right)}},\mathbf{A}} \right) = \sigma \left( {\mathbf{A}{\mathbf{H}^{\left( l \right)}}{\mathbf{W}^{\left( l \right)}}} \right)
\end{equation}
where, $\mathbf{H}^{(0)}$ is the eigenmatrix. When $\mathbf{D}$ is normalized, $\mathbf{A}$ benefits from a diagonal node degree matrix. $\mathbf{A}$ is the adjacency matrix preserving the graph's self-loop. Putting all this together, $\mathbf{A}$ is as follows:
\begin{equation}
\mathbf{A} = {\hat \mathbf{D}^{ - \frac{1}{2}}}\hat \mathbf{A}{\hat\mathbf{D}^{ - \frac{1}{2}}}
\end{equation}

\subsection{Multi-head attention with a GAT}

Another neural network model, a GAT \cite{song2019session}, \cite{martin2021attention}, processes graphical structured data in the propagation using self-attention techniques. In more detail, the GAT calculates the node states based on the node's neighbors, where each node is assigned an attention factor, calculated using its coefficient of attention.
\begin{equation}
{\alpha _{ij}} = \frac{{\exp \left( {{\rm{LeakyReLU}}\left( {{{\vec \alpha }^T}\left[ {{\rm{\mathbf{W}}}{{\vec h}_i}\left\| {{\rm{\mathbf{W}}}{{\vec h}_j}} \right.} \right]} \right)} \right)}}{{\sum\limits_{k \in {N_i}} {\exp \left( {{\rm{LeakyReLU}}\left( {{{\vec \alpha }^T}\left[ {{\rm{\mathbf{W}}}{{\vec h}_i}\left\| {{\rm{\mathbf{W}}}{{\vec h}_j}} \right.} \right]} \right)} \right)} }}
\end{equation}
where $\alpha_{(ij)}$ is the attention coefficient from adjacent node ($i\rightarrow j$). By the weight matrix $\left[ {{\rm{\mathbf{W}}}{{\vec h}_i}\left\| {{\rm{\mathbf{W}}}{{\vec h}_j}} \right.} \right]$, ($\mathbf{W}\in\mathbb{R^{F\times F}}$), $\alpha$ as a weight vector. $T$ stands for transposition, $\|$ stands for series operation.

The final output feature can be expressed as follows by combining the normalized attention coefficient with the linear combination of features.
\begin{equation}
{{\vec h}_i}^{'} = \sigma \left( {\sum\limits_{j \in {N_i}} {{\alpha _{ij}}{\rm{\mathbf{W}}}{{\vec h}_j}} } \right)  %{\vec h}_i^'
\end{equation}

The characteristic functions can be concatenated using the $k$ independent attention layers in Equation (9) through the multi-head attention technique, which is an appropriate approach for the learning process:
\begin{equation}
{{\vec h}_i}^{'} = \sigma \left( {\frac{1}{K}\sum\limits_{k = 1}^K {\sum\limits_{j \in {N_i}} {\alpha _{ij}^k{{\rm{\mathbf{W}}}^k}{{\vec h}_j}} } } \right)
\end{equation}

\subsection{Modeling the social influence of COVID-19}

Through deep learning, DeepInf can automatically identify hidden patterns at the user level and predict their influence. PPNP is a personalized neural prediction propagation algorithm. In this subsection, we propose a new DeepInf neural network model for deep learning that merges DeepInf, PPNP and APPNP. More specifically, it combines the personalized page ranking algorithms with the DeepInf architecture. The GAT/GCN networks are replaced with a personalized PageRank network model, and a novel algorithm called DeepPP is devised that combines PPNP and APPNP.

To adjust the size of the neighborhood influence, the transfer probability $\alpha$ is used in both the PPNP and APPNP methods. Therefore, the regression equation of a stealthy transport vector enhancement is combined with the prediction function $\mathbf{H}$ and the stealthy transport probability $\alpha$ in Equation (11):

\begin{equation}
{Z_{DeepPP}} = softmax\left( {\alpha {{\left( {{\mathbf{I}_n} - \left( {1 - \alpha } \right)\hat \mathbf{A}} \right)}^{ - 1}}\mathbf{H} + \alpha \mathbf{H}} \right)
\end{equation}

PPNP's linearity is reduced by adding $\alpha \mathbf{H}$. The flexibility of $\alpha \mathbf{H}$ is apparent in practical applications.

\begin{figure}[htb]
\centering
\subfigure[]{
\label{Fig3:sub:1}
\includegraphics[width=0.8\textwidth]{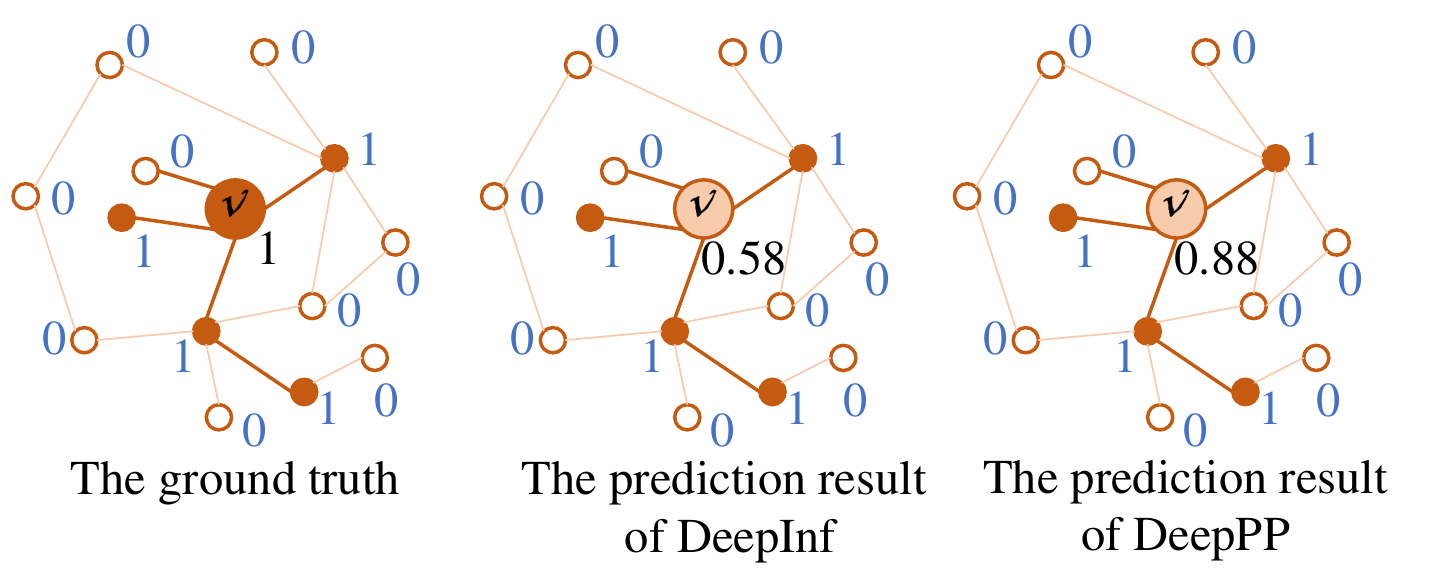}}
\subfigure[]{
\label{Fig3:sub:2}
\includegraphics[width=0.8\textwidth]{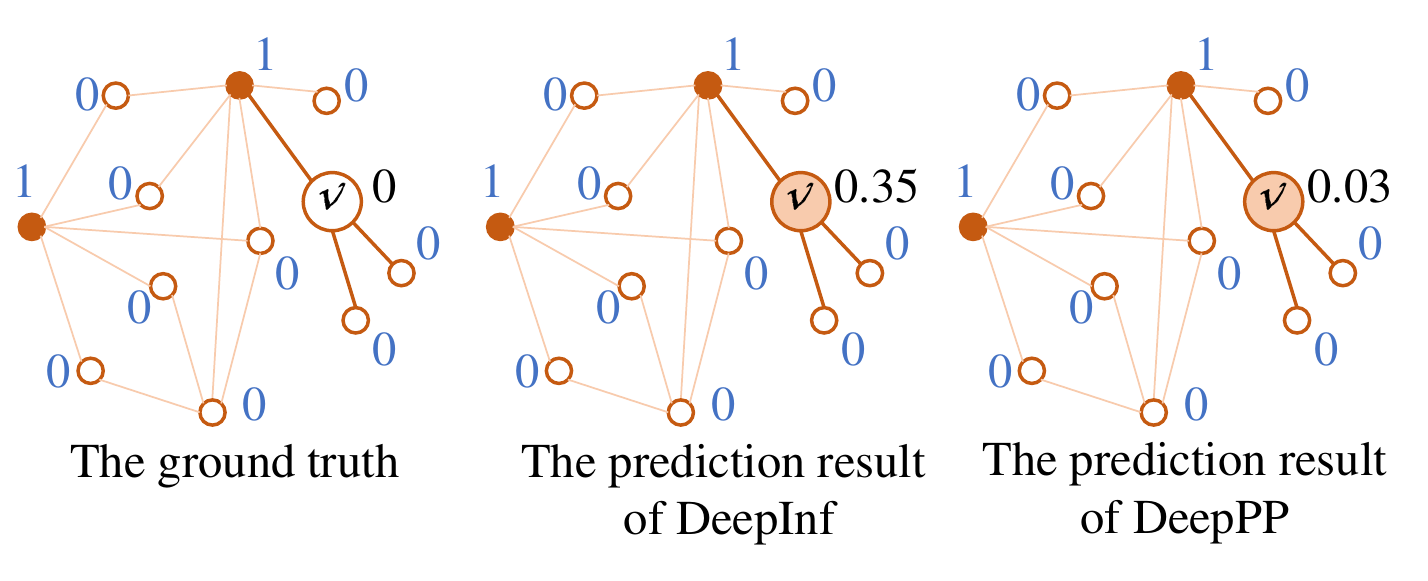}}
\caption{Two samples from the Digg dataset. (a) Social influence prediction with an active node. (b) Social influence prediction with an inactive node.}
\label{fig:3}
\end{figure}

To demonstrate the algorithm, Figure \ref{fig:3} shows two examples from the Digg dataset. The goal of social influence prediction is to predict the behavior state of target node $v$ from neighboring nodes. User $v$ represents the target node. Filled nodes represent social behavior $1$, and hollow nodes represent social behavior $0$. Figure \ref{Fig3:sub:1} shows a social influence prediction with an active node, while Figure \ref{Fig3:sub:2} shows a prediction with an inactive node. Our experiments show that DeepPP predicts the ground truth more accurately than DeepInf.

\section{Experiments}

In our experiments, we compared the DeepPP model with the conventional models GNN* \cite{4700287}, PPNP and APPNP \cite{klicpera2018personalized}, as well as the state-of-the-art models DeepInf \cite{qiu2018deepinf} and DeepEmLAN \cite{ZHAO2021382}. To avoid confusion with graph neural networks in a broader sense, the particular method we used has been named GNN*. GNN* is an extension to a common framework that includes information diffusion, relaxation mechanisms, and a random walk model. Its input can be a cyclic graph, a directed graph, an undirected graph, or a mixed graph. DeepEmLAN smoothly projects different types of attributes and topologies into the same semantic space, while retaining different types of attributes and topologies to the extent possible. The learning rate for all baselines was set to 0.1, and 1000 epochs for mean square error loss.

\subsection{Datasets}

To train the classification models, three categories of features were considered for ego-user:(a) vertex features; (b) pretrained network embeddings (in DeepWalk \cite{perozzi2014deepwalk}); and (c) hand-crafted features. They are listed in Table \ref{table:1}.

\begin{table}[htbp] %Â¿ÂªÃÂ¼ÃÂ»Â¸Ã¶Â±Ã­Â¸Ã±environmentÂ£Â¬Â±Ã­Â¸Ã±ÂµÃÃÂ»ÃÃÃÃh,hereÂ¡Â£
\centering
\caption{List of features for ego-user} %ÃÃÃÂ¾Â±Ã­Â¸Ã±ÂµÃÂ±ÃªÃÃ¢
\begin{tabular}{p{1.5cm}|p{6.6cm}} %ÃÃ¨ÃÃÃÃÃÂ¿ÃÂ»ÃÃÂµÃÂ¿Ã­Â¶ÃÂ£Â¬ÃÂ¿ÃÃÃÂªÂ»Â»Â¡Â£
\hline
\hline
Feature & Description \\ %ÃÃ&ÃÂ´Â·ÃÂ¸Ã´ÂµÂ¥ÃÂªÂ¸Ã±ÂµÃÃÃÃÃ \\Â±Ã­ÃÂ¾Â½Ã¸ÃÃ«ÃÃÃÂ»ÃÃ
\hline %Â»Â­ÃÂ»Â¸Ã¶ÂºÃ¡ÃÃÂ£Â¬ÃÃÃÃ¦ÂµÃÂ¾ÃÂ¶Â¼ÃÃÃÂ»ÃÃ¹ÃÃÂ£Â¬ÃÃ¢ÃÃ¯ÃÂ»Â¹Â²ÃÃ4ÃÃÃÃÃÃ
Vertex & PageRank \cite{klicpera2018personalized}, eigenvector centrality \cite{matsubara2012rise}, reciprocal of ego user's degree \cite{qiu2018deepinf}, coreness \cite{batagelj2001cores}, clustering coefficient \cite{schank2005approximating}, hub score and authority score \cite{zhang2015influenced}.\\
\hline
Embedding & Pre-trained network embedding in DeepWalk\\
\hline
Ego & The number/ratio of active neighbors \cite{LI2020107206}, the density of subgnetwork induced by active neighbors, and the number of connected components formed by active neighbors \cite{ugander2012structural}\\
\hline
\hline
\end{tabular}
\label{table:1}
\end{table}

\subsubsection{Four social networks in different domains}

We conducted the experiment with four datasets from different fields. Table \ref{table:2} presents the statistical information. $|V|$ and $|E|$ represent the number of nodes and the number of edges in graph $\mathcal{G}=(V, E)$, respectively. $N$ represents the number of observable instances.

\begin{table}[htbp]
	\centering
	\caption{Statistics for the datasets}
	\begin{tabular}{ccccc}
		\toprule  % Â¶Â¥Â²Â¿ÃÃ
		symbol & OAG & Digg & Twitter & Weibo\\
		\midrule  % ÃÃÂ²Â¿ÃÃ
		$|V|$ & 953,675 & 71,367 & 456,626 & 1,776,950\\
        $|E|$ & 4,151,463 & 1,731,658 & 12,508,413 & 308,489,739\\
        $N$ & 499,848 & 27,323 & 499,160 & 779,164\\
		\bottomrule  % ÂµÃÂ²Â¿ÃÃ
	\end{tabular}
\label{table:2}
\end{table}

\textbf{OAG}. The dataset links two large academic graphs, the Microsoft Academic Graph and Aminer \cite{10.1145/2835776.2835849}. We chose 20 major conferences in the areas of computer science and artificial intelligence, such as SIGCOMM,  SIGMOD, AAAI, NeurIPS, similar to the approach described in \cite{zhuang2018dual}. The social networks were defined as co-author networks, and social behavior was defined as citation behavior where one researcher cites a paper presented at the conference. What we are interested in is how a person's citation behavior is influenced by his/her collaborators.

\textbf{Digg}, a social news site, is based in the United States. There are two features, digging and burying, based on whether people agree with the story. In this dataset, we provide data on the stories that appeared on the front page of major newspapers in 2019.

\textbf{Twitter}. The Twitter dataset was built from propagations of the announcement of the discovery of the elusive Higgs boson on Twitter in July 2012. This is a Twitter friendship network where social behavior is mapped by retweets.

\textbf{Weibo} \cite{qiu2018deepinf}. Sina Weibo is the most popular microblogging service in China. It's a social media platform based on user relationships. The number of daily active users is hundreds of millions. This dataset contains target tracking networks and tweets from September 28, 2012 through to October 29, 2012.

\subsubsection{COVID-19 Datasets}
\begin{figure}[ht]
\begin{center}
\includegraphics*[width=0.8\textwidth]{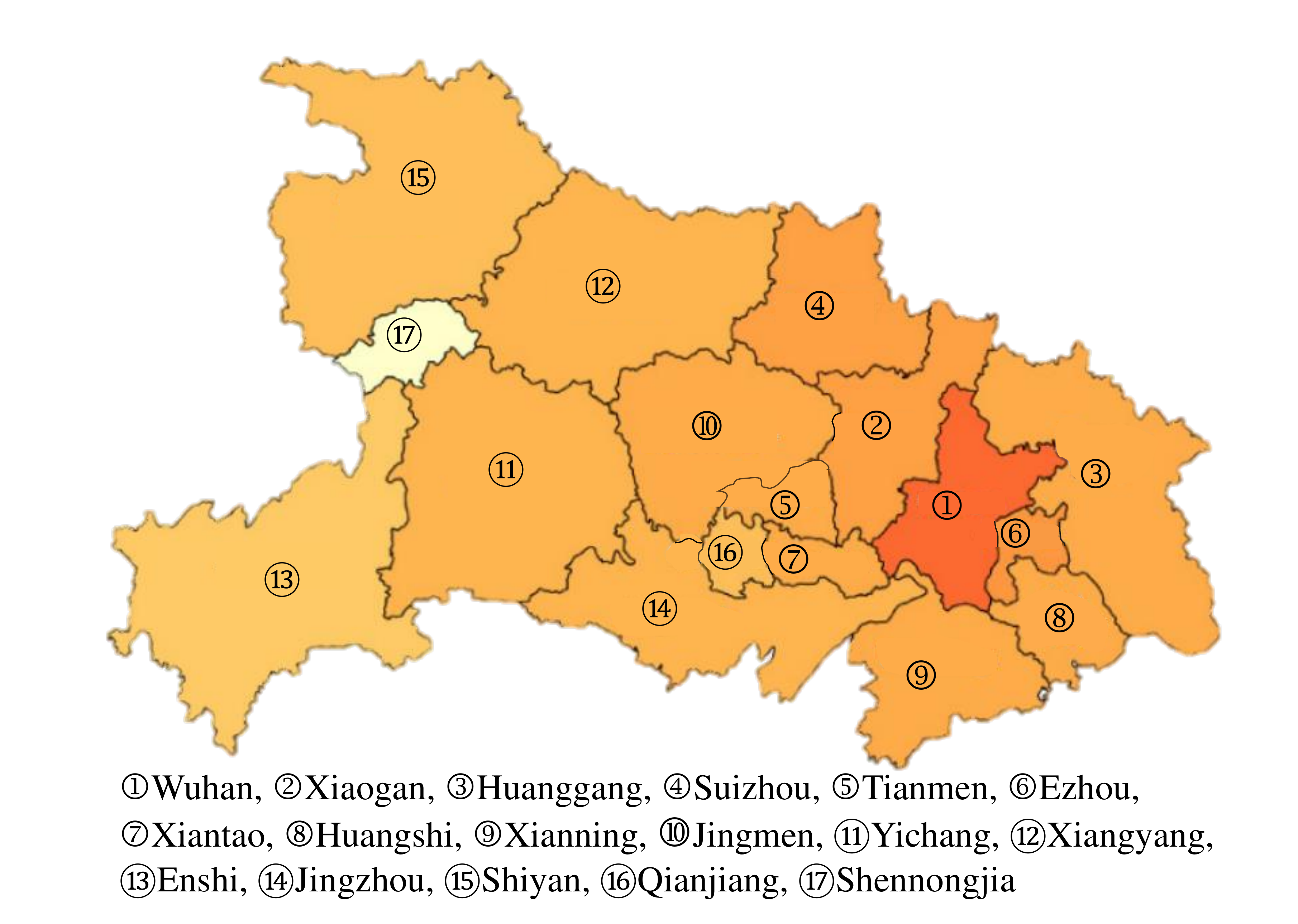}
\caption{Infection during COVID-19 in Hubei, China.}
\label{fig:4}
\end{center}
\end{figure}

\textbf{Hubei} (see Figure \ref{fig:4}). This is a dataset of infected cases provided by the Hubei Provincial Health Committee \footnote{http://wjw.hubei.gov.cn/fbjd}  \cite{2020Wuhan}. In December 2019, COVID-19 first appeared in Wuhan, the capital of Hubei Province, China. On January 21st, 2020, Hubei Provincial Health Committee reported the first case outside Wuhan. Then from February 15, 2020, the diagnostic policy was changed , resulting in a sharp increase in the number of recorded infection cases. Therefore, this data set is limited to the period from January 21 to February 14, 2020.

\begin{figure}[ht]
\begin{center}
\includegraphics*[width=0.8\textwidth]{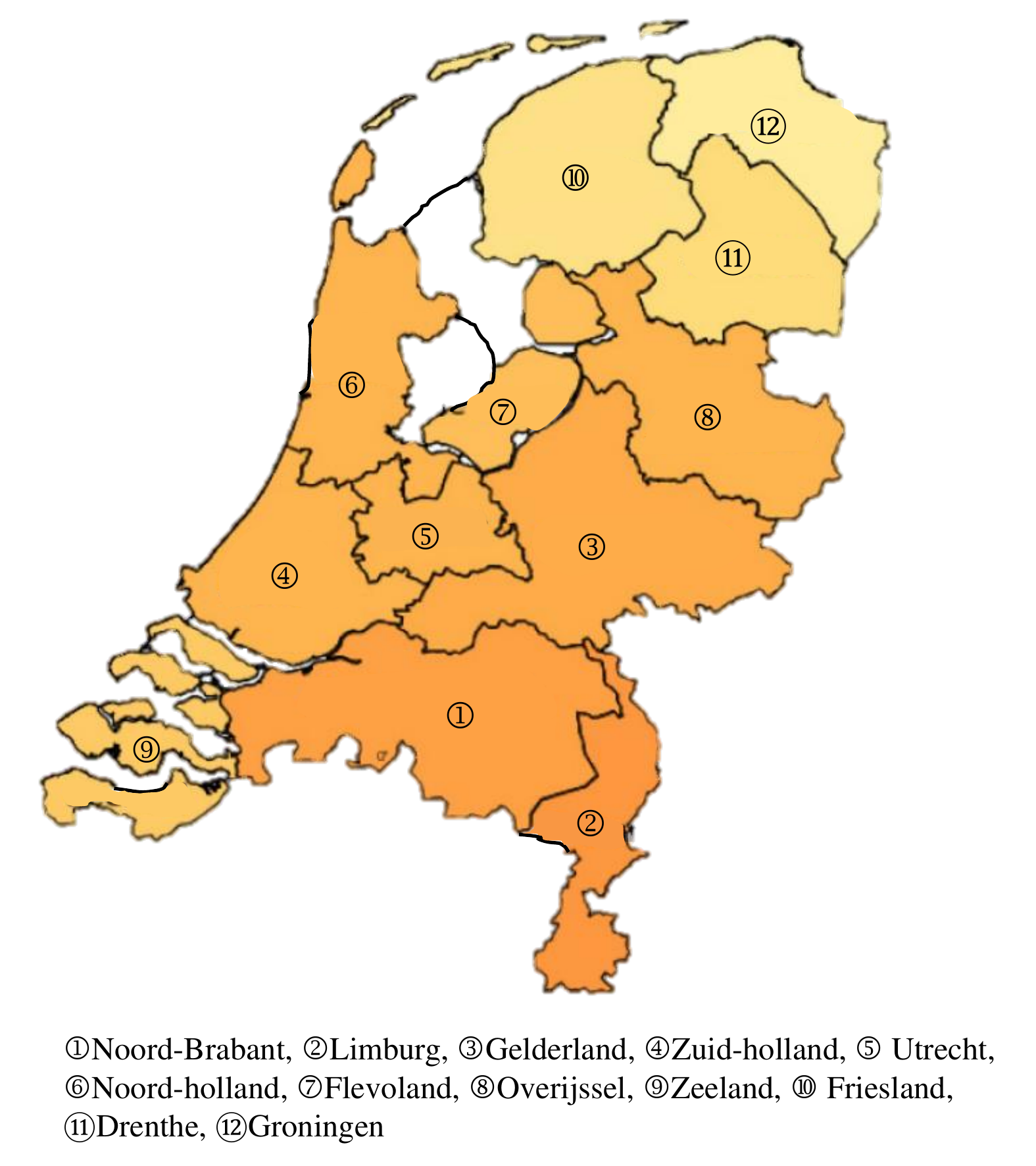}
\caption{Infections during COVID-19 in Holland.}
\label{fig:5}
\end{center}
\end{figure}

\textbf{Holland} (see Figure \ref{fig:5}). This dataset contains data on infection cases collected by the Dutch National Institute for Public Health and the Environment \footnote{https://www.rivm.nl} \cite{2021Maternal}. The first infection, diagnosed on February 27, 2020, went to Italy a week ago, after which the number of cases grew rapidly. Reported cases reached their peak at the end of March, followed by a downward trend in the number of daily reported cases. As the number of reported cases in Holland increased more slowly than that in Hubei, the overall infection period was longer and there are more data points. The gradually increasing number of infected cases is conducive to the accuracy of prediction.

\subsection{Evaluation}

The following metrics were used to measure performance in the experiments: area under curve (AUC), Precision (Prec.), Recall (Rec.), and $F1-Measure$ ($F1$) \cite{Wu_IJCAT}.

\begin{figure}[ht]
\begin{center}
\includegraphics*[width=\textwidth]{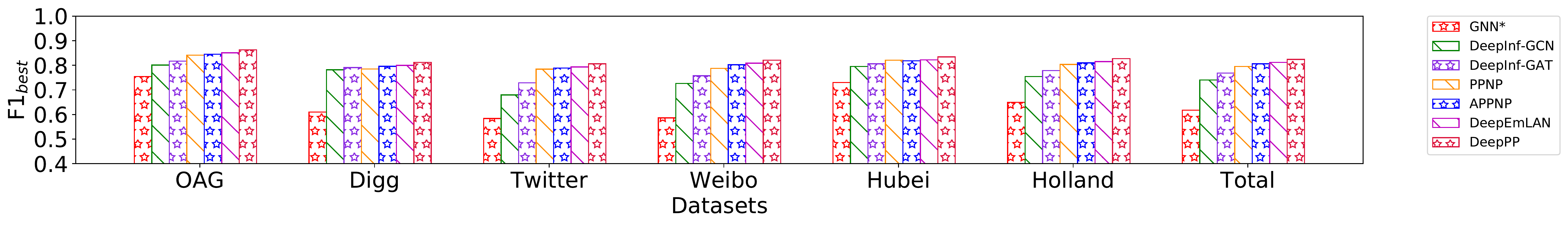}
\caption{A comparative analysis of DeepPP and baseline methods in F1$_{best}$ across six datasets.}
\label{fig:6}
\end{center}
\end{figure}

We first analyzed the overall performance of methods. We searched each parameter in the threshold space across the six datasets and calculated the corresponding $F1$ to get the most appropriate value, i.e., the best $F1-Measure$ ($F1_{best}$). Figure \ref{fig:6} shows the $F1_{best}$ score and the total average of the datasets. On $F1_{best}$, DeepPP was on average  $2.01\%$ higher than DeepEmLAN, $2.80\%$ higher than APPNP, $3.78\%$ higher than PPNP, $6.26\%$ higher than DeepInfGAT, and $8.73\%$ higher than DeepInf-GCN. Hence, DeepPP yielded superior results to the state-of-the-art methods.

To determine the effects of the hyper-parameters, we adjusted the transfer probability $\alpha$ from 0.2$\leq\alpha\leq$1.0, in 0.2-step increments. We then ran 1024 mini batches over 1000 epochs. Moreover, we used dropout technology with a dropout rate of 0.2. We compared predictions at $\alpha=$ 0.2, 0.4, 0.6, and 0.8. These results, as detailed in Table \ref{table:3}, clearly indicate that DeepPP outperformed the other six method in terms of AUC, Precision, Recall, and $F1\_Measure$. Of the two DeepInf algorithms \cite{qiu2018deepinf}, GAT had the highest predictive performance. According to Table \ref{table:3}, DeepPP, DeepEmLAN, APPNP and PPNP were competitive. However, DeepInf-GAT was especially good with regard to Precision on the Weibo, Hubei, and Holland datasets.

\begin{table}[!htbp]
%\footnotesize
\centering
%\begin{threeparttable}[b]   %ÃÂªÃÃÃÃÂ±Ã­Â¸Ã±ÃÃÃÃ¦ÃÃÃÂ¾Â½ÃÃÂ¢
\caption{Comparison of different methods with respect to AUC, Precision, Recall, and $F1-Measure$, at varying values for $\alpha$.}
\scalebox{0.6}{
\begin{tabular}{|c|c|c|c|c|c|} %Â±Ã­Â¸Ã±6ÃÃ ÃÂ«Â²Â¿Â¾ÃÃÃÃÃÃÂ¾
\hline
%\multicolumn{6}{|c|}{event}\\  %ÂºÃ¡ÃÃ²ÂºÃÂ²Â¢6ÃÃÂµÂ¥ÃÂªÂ¸Ã±  ÃÂ½Â²Ã ÃÃ­Â¼ÃÃÃºÃÃ
\multirow{2}*{Data}&\multirow{2}*{Model}&$\alpha$=0.2&$\alpha$=0.4&$\alpha$=0.6&$\alpha$=0.8\\  %ÃÃÃÃ²ÂºÃÂ²Â¢6ÃÃÂµÂ¥ÃÂªÂ¸Ã±
\cline{3-6} & &AUC~~~Prec.~~~Rec.~~~F1&AUC~~~Prec.~~~Rec.~~~F1&AUC~~~Prec.~~~Rec.~~~F1&AUC~~~Prec.~~~Rec.~~~F1\\
\hline
\multirow{7}*{OAG}&GNN*\cite{4700287}&62.99 28.16 76.56 43.02&63.60 30.02 74.25 42.09&64.49 31.07 71.74 42.86&65.40 30.08 60.73 43.51\\  %ÃÃÃÃ²ÂºÃÂ²Â¢6 ÃÃÂµÂ¥ÃÂªÂ¸Ã±
\cline{2-6}  %ÃÂªÂµÃÂ¶Ã¾ÃÃÂµÂ½ÂµÃ6ÃÃÃÃ­Â¼ÃÂºÃ¡ÃÃ
&DeepInf-GCN\cite{qiu2018deepinf}&60.74 28.16 71.86 40.54&61.65 28.36 72.74 41.65&62.76 29.05 73.28 42.31&63.54 30.30 74.37 43.01\\
\cline{2-6}  %ÃÂªÂµÃÂ¶Ã¾ÃÃÂµÂ½ÂµÃ6ÃÃÃÃ­Â¼ÃÂºÃ¡ÃÃ
&DeepInf-GAT\cite{qiu2018deepinf}&68.51 38.36 58.88 46.00&69.58 39.24 59.05 46.97&70.66 40.00 59.99 47.75&71.80 40.79 60.97 48.87\\
\cline{2-6}  %ÃÂªÂµÃÂ¶Ã¾ÃÃÂµÂ½ÂµÃ6ÃÃÃÃ­Â¼ÃÂºÃ¡ÃÃ
&PPNP\cite{klicpera2018personalized}&63.17 29.97 76.02 42.97&64.15 30.61 73.79 43.22&65.07 31.62 71.25 43.62&65.90 33.48 64.59 44.19\\
\cline{2-6}
&APPNP\cite{klicpera2018personalized}& 64.94 30.91 73.87 43.64&65.47 32.42 69.31 44.02&65.78 33.03 66.35 44.18&65.96 33.36 65.51 44.23\\
\cline{2-6}
&DeepEmLAN\cite{ZHAO2021382}&64.01 31.15 69.94 42.59&64.61 32.24 66.43 43.17&65.00 32.56 66.32 44.16&65.62 33.02 65.86 44.29\\
\cline{2-6}
&DeepPP&63.63 30.62 73.16 43.06&65.12 32.50 68.70 43.80&66.24 33.21 66.70 44.37&66.05 33.31 66.08 44.36\\
\hline

\multirow{7}*{Digg}&GNN*\cite{4700287}&83.90 51.85 70.26 62.87&84.19 61.18 69.32 61.47&84.28 65.36 68.09 64.16&84.81 65.46 69.61 70.07\\  %ÃÃÃÃ²ÂºÃÂ²Â¢6ÃÃÂµÂ¥ÃÂªÂ¸Ã±
\cline{2-6}  %ÃÂªÂµÃÂ¶Ã¾ÃÃÂµÂ½ÂµÃ6ÃÃÃÃ­Â¼ÃÂºÃ¡ÃÃ
&DeepInf-GCN\cite{qiu2018deepinf}&81.06 56.28 64.44 59.36&82.18 57.15 65.65 60.22&83.27 58.04 66.54 61.76&84.16 58.73 67.62 62.89\\
\cline{2-6}  %ÃÂªÂµÃÂ¶Ã¾ÃÃÂµÂ½ÂµÃ6ÃÃÃÃ­Â¼ÃÂºÃ¡ÃÃ
&DeepInf-GAT\cite{qiu2018deepinf}&86.32 63.59 67.38 69.23&87.67 64.31 68.64 70.15&88.59 65.24 69.51 71.08&90.17 66.80 70.48 72.19\\
\cline{2-6}  %ÃÂªÂµÃÂ¶Ã¾ÃÃÂµÂ½ÂµÃ6ÃÃÃÃ­Â¼ÃÂºÃ¡ÃÃ
&PPNP\cite{klicpera2018personalized}&81.92 52.41 68.33 59.02&85.23 61.75 66.87 63.53&87.51 66.26 67.91 67.06&89.05 68.32 71.53 69.78\\
\cline{2-6}
&APPNP\cite{klicpera2018personalized}&85.77 59.50 71.19 64.36&86.81 63.25 68.75 65.52&87.92 68.70 65.25 66.59&89.08 68.39 71.43 69.36\\
\cline{2-6}
&DeepEmLAN\cite{ZHAO2021382}&83.38 56.74 70.49 62.16&86.25 62.36 69.41 65.44&88.10 67.12 71.85 68.76&89.26 70.03 72.46 70.05\\
\cline{2-6}
&DeepPP&84.89 57.06 71.90 63.52&87.47 63.55 71.25 66.89&89.17 69.24 71.19 69.98&90.64 72.35 71.15 71.38\\
\hline

\multirow{7}*{Twitter}&GNN*\cite{4700287}&74.26 41.76 68.31 50.15&74.90 43.09 67.59 50.94&75.81 44.48 66.72 52.03&76.92 44.39 66.00 53.13\\  %ÃÃÃÃ²ÂºÃÂ²Â¢6ÃÃÂµÂ¥ÃÂªÂ¸Ã±
\cline{2-6}  %ÃÂªÂµÃÂ¶Ã¾ÃÃÂµÂ½ÂµÃ6ÃÃÃÃ­Â¼ÃÂºÃ¡ÃÃ
&DeepInf-GCN\cite{qiu2018deepinf}&74.04 45.16 64.04 50.35&74.93 46.02 65.23 51.16&75.84 46.24 66.00 52.18&76.61 44.41 66.75 53.27\\
\cline{2-6}  %ÃÂªÂµÃÂ¶Ã¾ÃÃÂµÂ½ÂµÃ6ÃÃÃÃ­Â¼ÃÂºÃ¡ÃÃ
&DeepInf-GAT\cite{qiu2018deepinf}&75.11 45.45 65.25 54.21&76.20 46.15 67.47 55.13&76.86 46.77 68.36 56.22&77.21 47.43 69.18 56.92\\
\cline{2-6}  %ÃÂªÂµÃÂ¶Ã¾ÃÃÂµÂ½ÂµÃ6ÃÃÃÃ­Â¼ÃÂºÃ¡ÃÃ
&PPNP\cite{klicpera2018personalized}&73.98 42.25 64.59 51.01&75.02 42.51 66.83 52.02&76.53 44.61 66.58 53.39&77.32 45.52 66.58 54.71\\
\cline{2-6}
&APPNP\cite{klicpera2018personalized}&75.97 43.48 67.28 52.76&76.36 44.62 66.25 53.35&77.22 46.32 65.26 54.20&77.69 46.47 65.26 54.84\\
\cline{2-6}
&DeepEmLAN\cite{ZHAO2021382}&73.28 42.09 68.32 52.61&74.19 43.36 67.13 53.76&75.48 44.09 68.91 54.88&76.09 45.62 68.54 55.56\\
\cline{2-6}
&DeepPP&74.55 41.75 67.27 51.53&76.38 45.17 65.57 53.46&77.74 46.43 67.02 55.02&78.66 47.73 66.73 55.61\\
\hline

\multirow{7}*{Weibo}&GNN*\cite{4700287}&76.36 45.04 64.99 53.15&77.01 43.89 64.18 52.74&77.86 45.93 65.56 53.88&78.82 47.06 66.18 55.00\\  %ÃÃÃÃ²ÂºÃÂ²Â¢6ÃÃÂµÂ¥ÃÂªÂ¸Ã±
\cline{2-6}  %ÃÂªÂµÃÂ¶Ã¾ÃÃÂµÂ½ÂµÃ6ÃÃÃÃ­Â¼ÃÂºÃ¡ÃÃ
&DeepInf-GCN\cite{qiu2018deepinf}&76.01 45.66 64.14 53.57&77.17 46.73 64.02 53.85&78.21 47.24 65.14 54.63&78.99 47.52 66.81 55.05\\
\cline{2-6}  %ÃÂªÂµÃÂ¶Ã¾ÃÃÂµÂ½ÂµÃ6ÃÃÃÃ­Â¼ÃÂºÃ¡ÃÃ
&DeepInf-GAT\cite{qiu2018deepinf}&77.82 46.08 64.56 55.39&78.26 46.96 64.23 56.26&80.34 48.21 65.18 56.75&81.52 48.99 68.11 57.42\\
\cline{2-6}  %ÃÂªÂµÃÂ¶Ã¾ÃÃÂµÂ½ÂµÃ6ÃÃÃÃ­Â¼ÃÂºÃ¡ÃÃ
&PPNP\cite{klicpera2018personalized}&75.98 45.89 64.34 53.76&77.25 46.81 64.18 55.21&78.49 47.78 65.03 55.89&80.09 48.68 67.51 56.49\\
\cline{2-6}
&APPNP\cite{klicpera2018personalized}&77.96 46.45 64.15 54.61&78.48 47.33 64.01 55.36&79.23 47.96 65.26 56.08&80.15 48.83 67.49 56.63\\
\cline{2-6}
&DeepEmLAN\cite{ZHAO2021382}&75.99 44.46 63.01 53.78&77.65 45.71 63.96 54.47&78.44 46.68 65.83 55.58&81.43 47.58 66.92 56.65\\
\cline{2-6}
&DeepPP&76.49 46.42 64.32 55.57&78.40 47.39 64.05 56.19&79.99 48.56 65.73 56.36&82.13 49.04 67.75 57.29\\
\hline

\multirow{7}*{Hubei}&GNN*\cite{4700287}&77.89 40.03 71.85 61.04&78.57 40.85 70.59 61.67&80.03 41.26 70.34 62.85&81.50 42.03 69.88 63.09\\  %ÃÃÃÃ²ÂºÃÂ²Â¢6ÃÃÂµÂ¥ÃÂªÂ¸Ã±
\cline{2-6}  %ÃÂªÂµÃÂ¶Ã¾ÃÃÂµÂ½ÂµÃ6ÃÃÃÃ­Â¼ÃÂºÃ¡ÃÃ
&DeepInf-GCN\cite{qiu2018deepinf}&79.68 40.02 64.64 62.60&80.01 40.95 65.47 63.56&80.67 41.48 66.38 64.37&81.05 42.51 67.26 65.42\\
\cline{2-6}  %ÃÂªÂµÃÂ¶Ã¾ÃÃÂµÂ½ÂµÃ6ÃÃÃÃ­Â¼ÃÂºÃ¡ÃÃ
&DeepInf-GAT\cite{qiu2018deepinf}&79.85 40.27 67.04 66.45&80.29 41.73 68.43 67.59&81.08 42.86 69.85 68.36&81.67 43.95 70.91 69.71\\
\cline{2-6}  %ÃÂªÂµÃÂ¶Ã¾ÃÃÂµÂ½ÂµÃ6ÃÃÃÃ­Â¼ÃÂºÃ¡ÃÃ
&PPNP\cite{klicpera2018personalized}&76.24 41.26 66.33 67.54&78.60 42.10 68.04 68.72&80.52 42.86 69.51 69.85&82.00 43.98 70.85 70.69\\
\cline{2-6}
&APPNP\cite{klicpera2018personalized}&79.49 42.58 68.72 68.23&79.99 43.46 69.25 69.08&80.31 43.77 70.81 70.26&82.21 44.59 71.72 71.05\\
\cline{2-6}
&DeepEmLAN\cite{ZHAO2021382}&78.05 42.13 73.86 68.72&79.14 43.27 73.06 69.98&80.46 44.13 72.59 70.59&81.85 44.83 71.69 71.08\\
\cline{2-6}
&DeepPP&78.90 43.76 69.03 68.16&80.57 44.14 71.02 69.95&82.18 44.53 71.66 70.47&83.01 45.76 72.56 71.58\\
\hline

\multirow{7}*{Holland}&GNN*\cite{4700287}&77.01 37.27 63.39 60.54&78.24 38.10 64.71 61.01&79.01 39.46 65.46 61.88&79.90 40.24 66.58 62.97\\  %ÃÃÃÃ²ÂºÃÂ²Â¢6ÃÃÂµÂ¥ÃÂªÂ¸Ã±
\cline{2-6}  %ÃÂªÂµÃÂ¶Ã¾ÃÃÂµÂ½ÂµÃ6ÃÃÃÃ­Â¼ÃÂºÃ¡ÃÃ
&DeepInf-GCN\cite{qiu2018deepinf}&77.44 37.12 64.85 61.32&78.25 37.59 65.74 62.00&79.09 38.78 66.62 62.77&79.91 39.34 67.45 63.48\\
\cline{2-6}  %ÃÂªÂµÃÂ¶Ã¾ÃÃÂµÂ½ÂµÃ6ÃÃÃÃ­Â¼ÃÂºÃ¡ÃÃ
&DeepInf-GAT\cite{qiu2018deepinf}&77.53 39.34 67.00 61.86&78.26 40.45 67.57 62.93&79.14 41.03 68.41 63.85&79.96 41.87 69.52 64.63\\
\cline{2-6}  %ÃÂªÂµÃÂ¶Ã¾ÃÃÂµÂ½ÂµÃ6ÃÃÃÃ­Â¼ÃÂºÃ¡ÃÃ
&PPNP\cite{klicpera2018personalized}&76.17 39.26 66.27 66.63&78.33 40.03 67.36 67.57&79.41 41.10 68.11 68.13&79.62 41.93 68.94 68.95\\
\cline{2-6}
&APPNP\cite{klicpera2018personalized}&77.75 40.01 66.32 66.74&78.71 40.74 67.49 68.25&79.80 41.68 69.01 69.68&79.84 42.71 69.86 70.36\\
\cline{2-6}
&DeepEmLAN\cite{ZHAO2021382}&77.84 41.15 72.30 66.48&78.25 41.93 71.99 68.31&79.04 42.72 71.28 69.66&79.79 42.92 70.17 70.13\\
\cline{2-6}
&DeepPP&77.26 40.52 64.36 66.91&78.78 41.76 66.43 68.65&80.18 42.44 68.31 69.74&80.39 43.38 70.11 70.59\\
\hline
\end{tabular}}
%\begin{tablenotes}  %Â±Ã­Â¸Ã±ÃÃÃÃ¦Â½ÃÃÂ¢Â¿ÂªÃÂ¼
%     \item[note:] GNN*\cite{4700287}, DeepInf-GCN and DeepInf-GAT\cite{qiu2018deepinf}, PPNP and APPNP\cite{klicpera2018personalized}, %DeepEmLAN\cite{ZHAO2021382}.
%  \end{tablenotes}
%  \end{threeparttable}
\label{table:3}
\end{table}

Some models based on deep learning were proposed to analyze the social influence, such as the famous DeepInf. We designed a novel model to improve the accuracy of prediction, which draws on DeepInf, PPNP and APPNP as inspiration. However, compared to the baseline methods, DeepPP has: more flexibility in practical applications, a more convenient way to adjust the parameters; and better performance with large and small datasets.

\subsection{Parameters analysis}

In terms of the gauging the influence of the other parameters, there are three categories to test: 1) the basic training parameters, which include the window size and the $z$-variable dimension of the potential space; 2) the temporal convolution network (TCN) unit parameters, i.e., the filter size and the TCN levels; and 3) the score attention parameters. For all these experiments, we set $\alpha$ to 0.8.

\begin{figure}[ht]
\begin{center}
\includegraphics*[width=\textwidth]{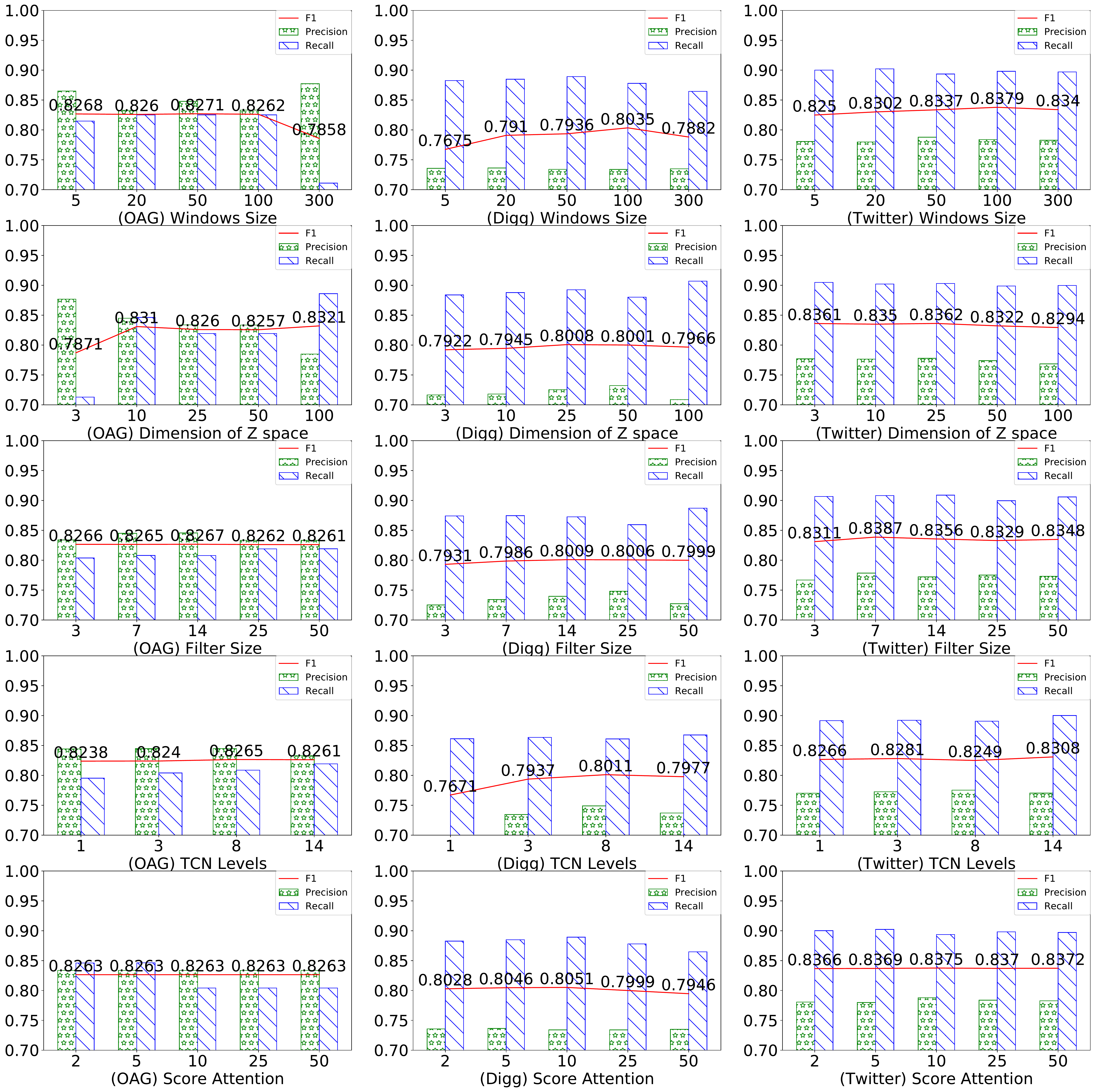}
\caption{The effect of parameters on OAG, Digg and Twitter}
\label{fig:7}
\end{center}
\end{figure}

\begin{figure}[ht]
\begin{center}
\includegraphics*[width=\textwidth]{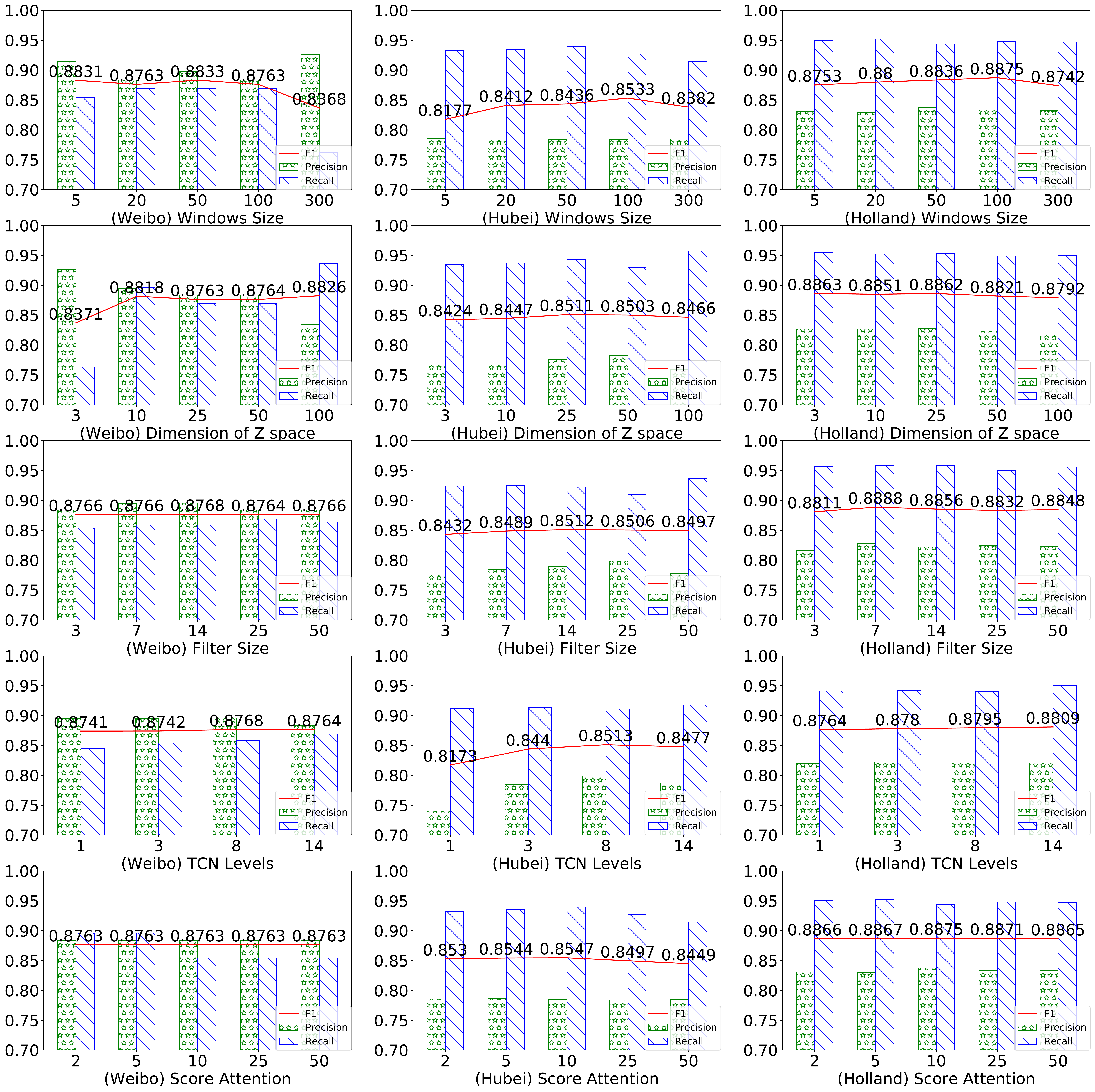}
\caption{The effect of parameters on Weibo, Hubei and Holland}
\label{fig:8}
\end{center}
\end{figure}

First, we studied the effects of changing the windows size. This directly affects the length of the time-dependencies in the historical data. The larger the window, the more data dependencies that can be captured. However, as the window size increases, so does the computing power required, which in turn affects the detection speed. The first row of Figures \ref{fig:7} and \ref{fig:8} shows that five window sizes were tested, i.e., 5, 20, 50, 100, 300. OAG and Weibo reached maximum $F1-score$ at a window size of 50, while Digg, Twitter, Hubei and Holland reached their maximums at a window size of 100. From this, we determined that the optimal window size relates to the composition of the dataset. A small window expresses better performance for datasets with weak time correlations. However, with time-dependent datasets, small windows cannot capture long-term dependencies. Additionally, observing the six datasets, all showed a performance degradation with a window size of 300. This indicates that, if the length of the data is too long, the generalization ability of the model will decrease. In addition, a too-large window will lead to a rapid increase in the required computing power, a larger model scale, and a slower training speed.

Next, we studied the effects of the second basic parameter, being the variable $z$ in the \textit{m}-dimension. The second row of Figures \ref{fig:7} and \ref{fig:8} shows the results for five \textit{z}-variable dimensions. What we see is poor performance with the OAG dataset with a small \textit{z}-dimension. This is because the dimension is mapped to a small potential variable space, which results in a large amount of information loss in the encoder stage. In turn, the decoder is unable to recover, resulting in performance degradation. We also observed that changes in the \textit{z}-dimension had little effect on performance with the Hubei and Holland datasets. As $z$ increased, the loss in the training process changed greatly. However, many iterations helped the models to stabilize.

The TCN filter size was next. Here, we fixed the expansion factor $d$, so only the filter size needed to be adjusted to change the field size. The third row of Figures \ref{fig:7} and \ref{fig:8} show the results for five filter sizes. As shown, the optimal filter size for the smaller Digg dataset was 7, while for the Hubei and Holland datasets, the optimal size was 14. This indicates that the optimal filter size is determined by the size of the dataset. However, due to the fixed expansion coefficient, the choice of filter size has little effect on the final performance of the model.

In terms of TCN levels, we found that changes in the TCN level directly affected the scale of the DeepPP model. The fourth row of Figures \ref{fig:7} and \ref{fig:8} shows the results for four TCN levels. From a data perspective, OAG and Digg (group training) have a smaller data scale, and so we saw better performance when using a smaller TCN level. Conversely, for the large-scale Hubei and Holland datasets, representation performance was greatly reduced at a TCN level of 1. Moreover, the smaller models could not capture the time dependencies effectively. From a model point of view, performance was excellent with a TCN level of 8. However, at a TCN level of 14, although the Weibo dataset showed a slight improvement, the size of the model almost doubled, which is unworkable in practice.

Next, we looked at score attention, varying the parameter's value across 2, 5, 10, 25, 50. The role of this mechanism is to improve the accuracy of abnormal data that is close to normal. The results, appearing in the fifth row of Figures \ref{fig:7} and \ref{fig:8}, show that varying this parameter has little effect on the results. This therefore warrants further attention in our ablation experiments.

\begin{figure}[!htbp]
\centering
\subfigure[]{
\label{Fig:sub:1}
\includegraphics[width=0.31\textwidth]{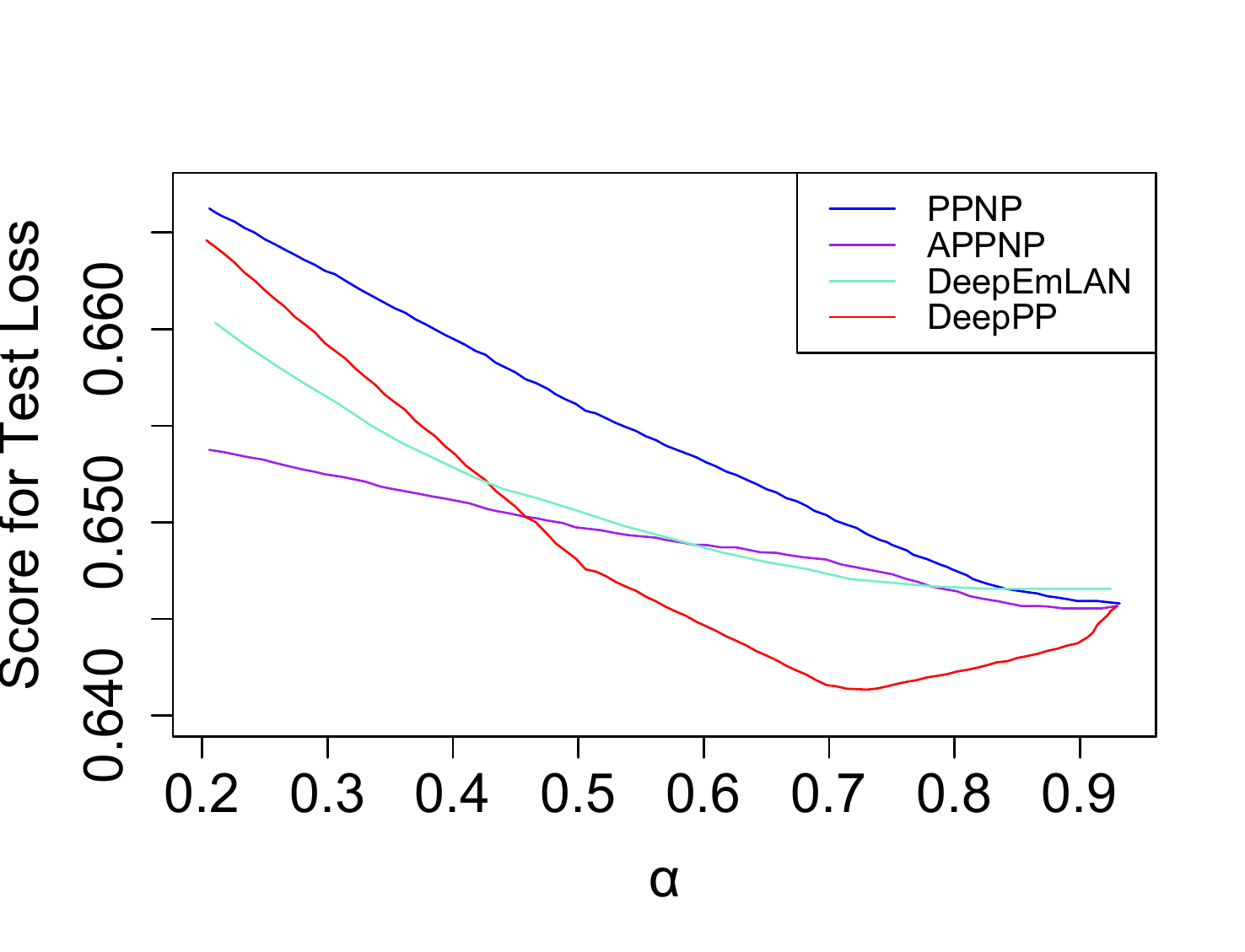}}
\subfigure[]{
\label{Fig:sub:2}
\includegraphics[width=0.31\textwidth]{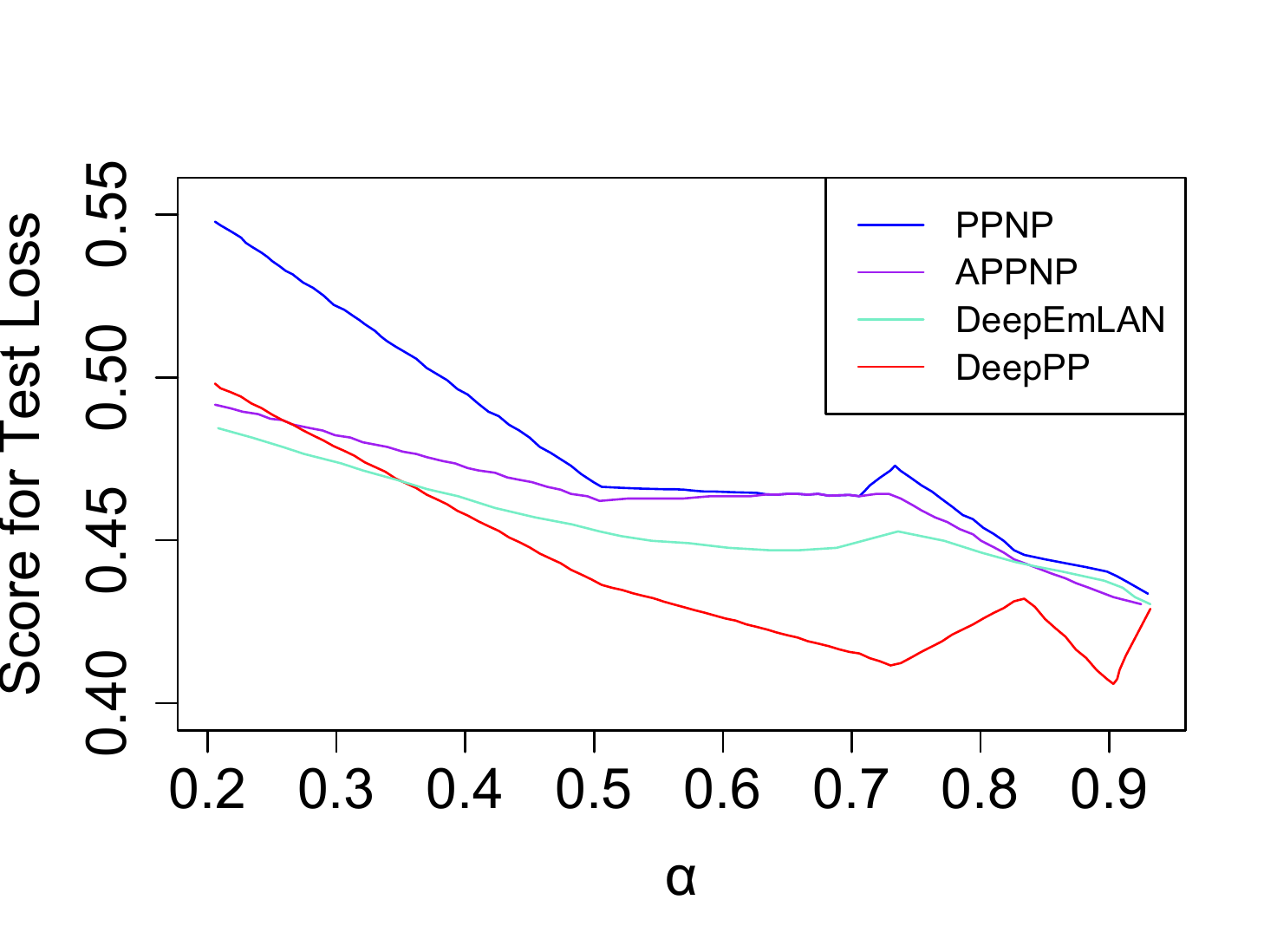}}
\subfigure[]{
\label{Fig:sub:3}
\includegraphics[width=0.31\textwidth]{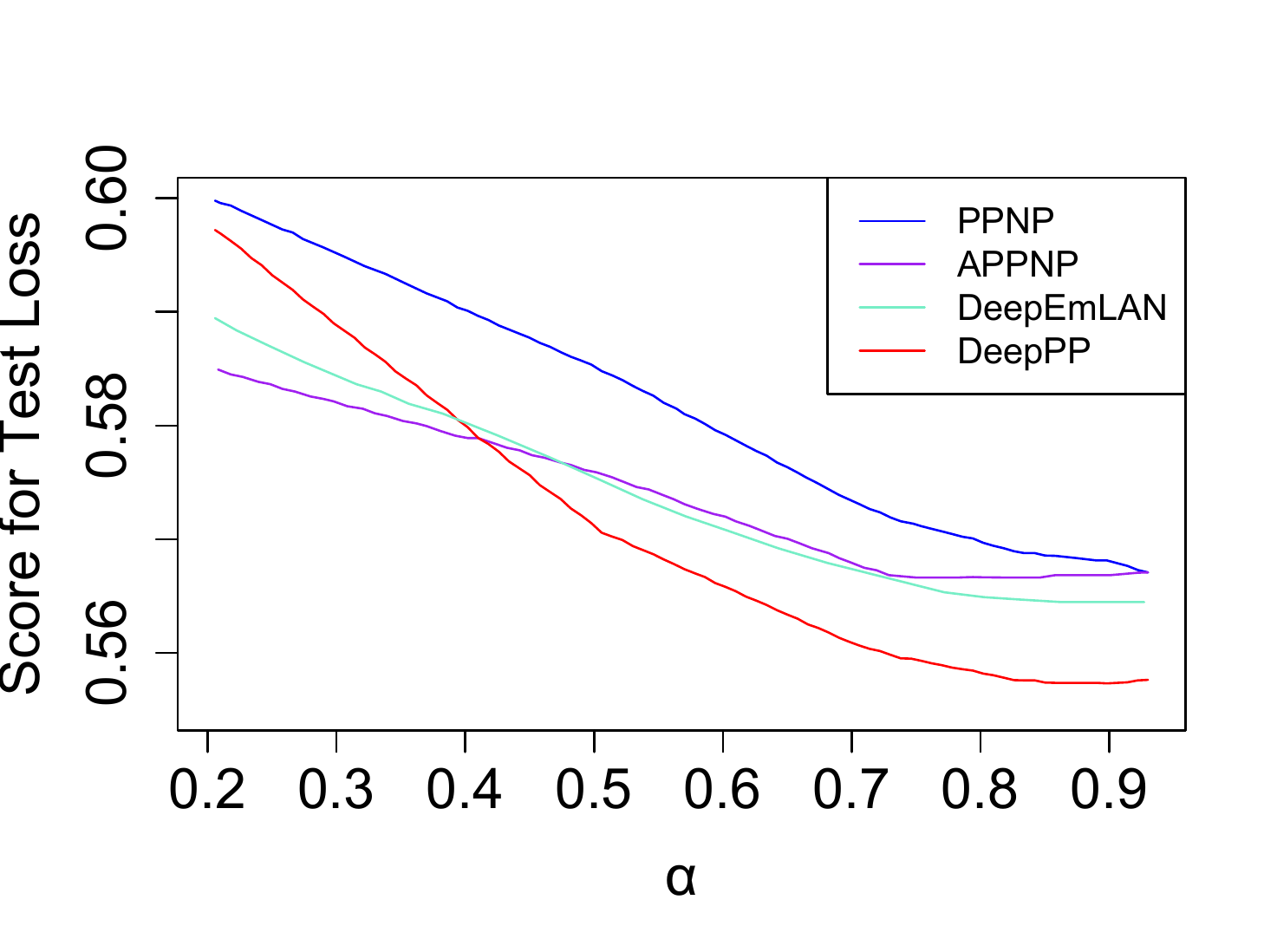}}
\subfigure[]{
\label{Fig:sub:4}
\includegraphics[width=0.31\textwidth]{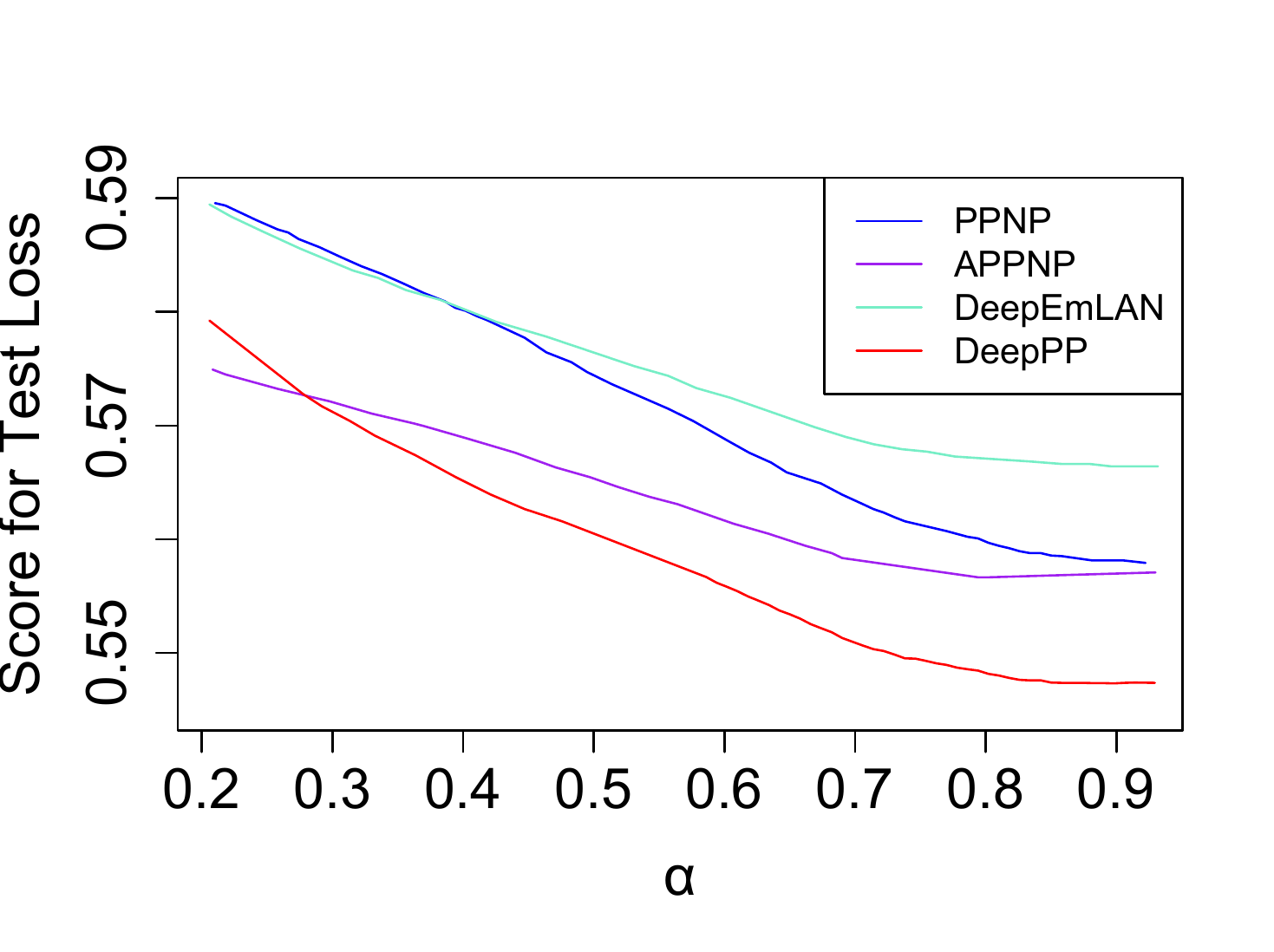}}
\subfigure[]{
\label{Fig:sub:5}
\includegraphics[width=0.31\textwidth]{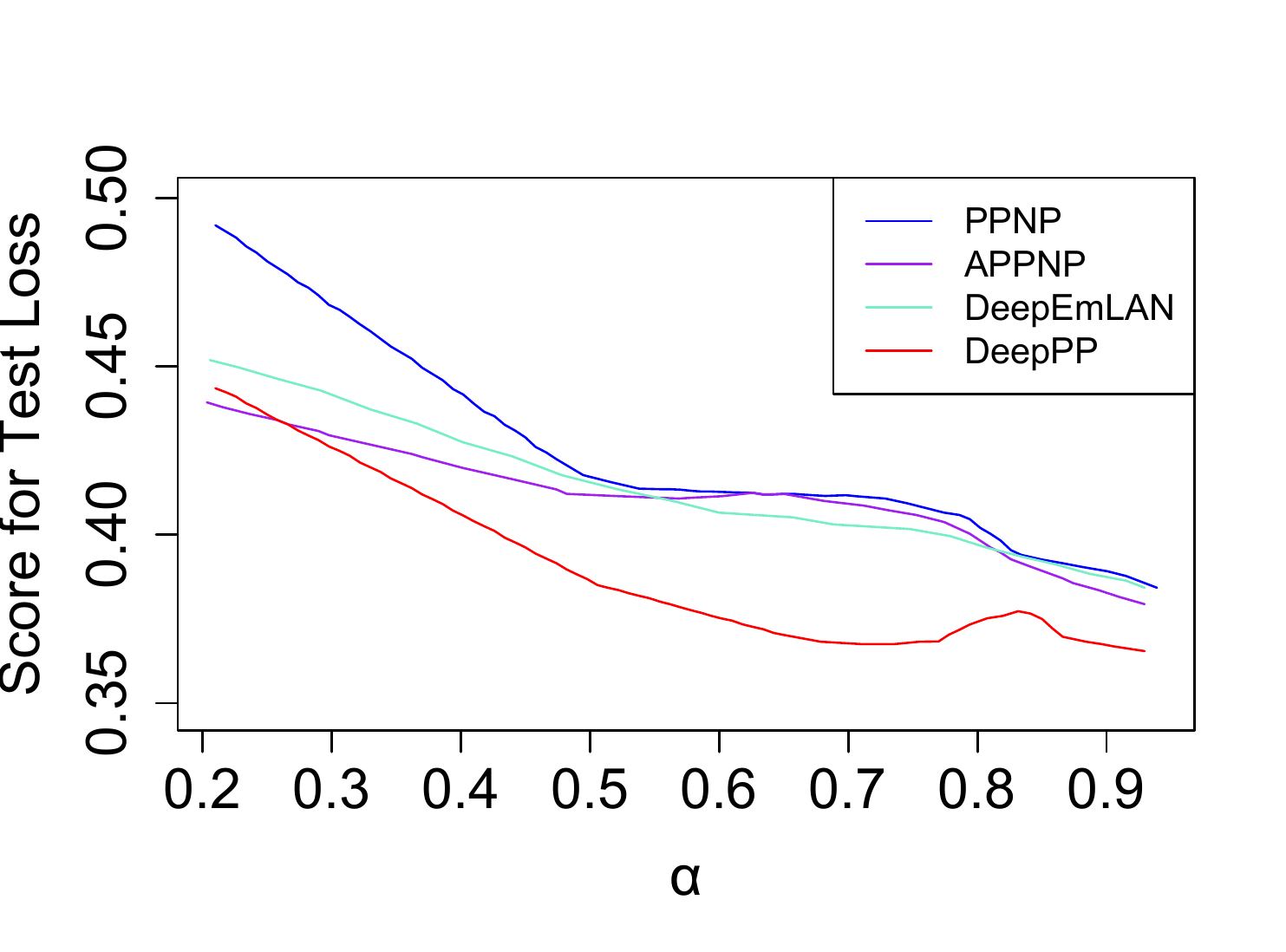}}
\subfigure[]{
\label{Fig:sub:6}
\includegraphics[width=0.31\textwidth]{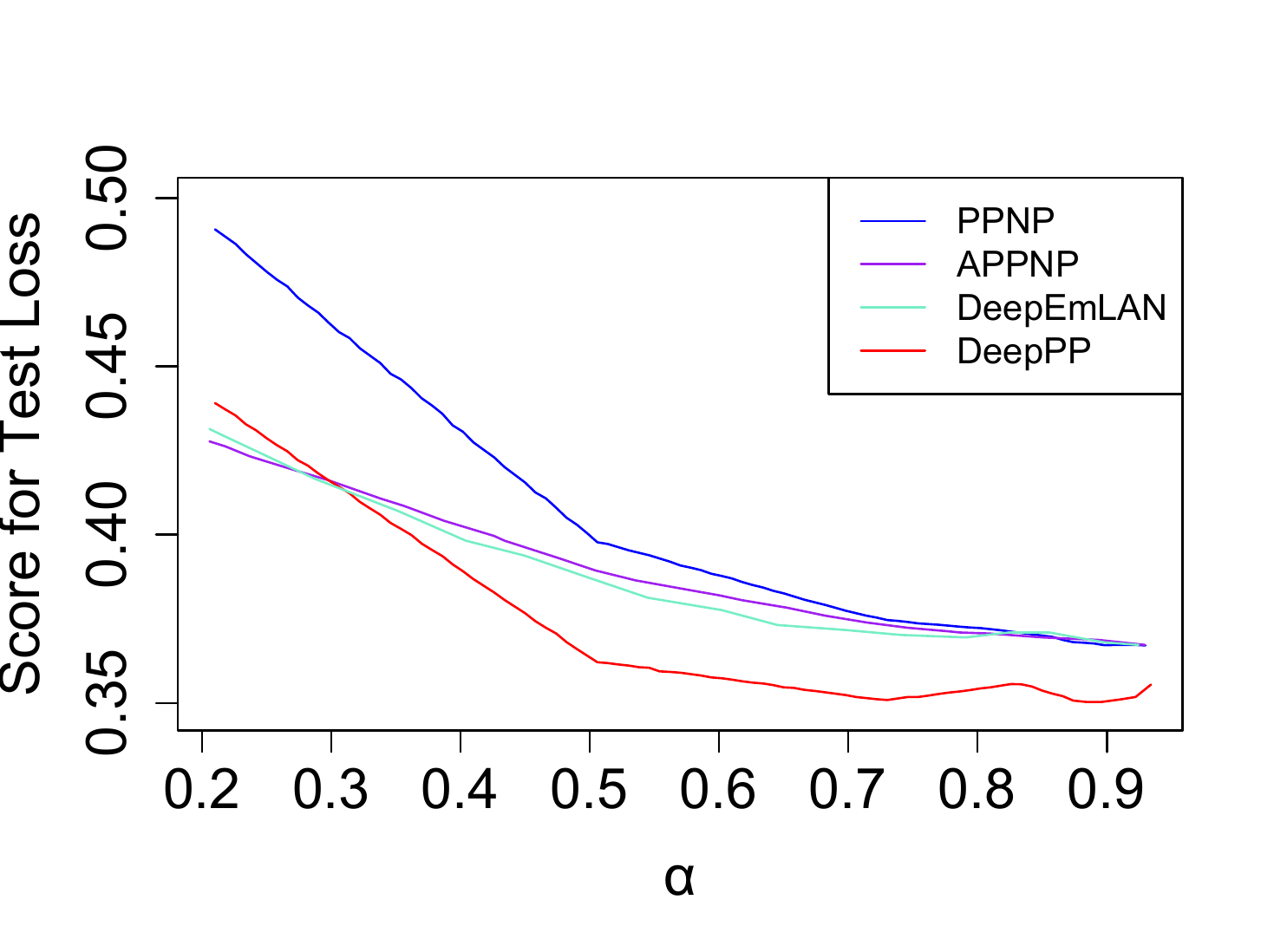}}
\caption{Test loss with $\alpha\in(0.2,1.0)$. (a) Test loss on OAG, (b) Test loss on Digg, (c) Test loss on Twitter, (d) Test loss on Weibo, (e) Test loss on COVID-19 in Hubei, (f) Test loss on COVID-19 in Holland.}
\label{Fig:9}
\end{figure}

\textbf{Test Loss}.  Figure \ref{Fig:9} shows a common trend, namely, that the final test loss decreases with an increase in $\alpha$. In fact, the results show that the larger $\alpha$ is, the less the test loss and the better the performance. From a comparison of four models with different datasets, as shown in Figure \ref{Fig:9}, we see that DeepPP had better performance than any of the other algorithms.

\subsection{Spread analysis of the COVID-19}

To illustrate the advantages and disadvantages of our method, we chose a COVID-19 dataset from two different regions, Hubei and Holland. Although these two areas do not represent complete representations of COVID-19's spread, they can reflect the spread of infectious diseases in general.

\subsubsection{Hubei}

\begin{figure}[!htbp]
\centering
\subfigure[]{
\label{Fig10:sub:1}
\includegraphics[width=0.31\textwidth]{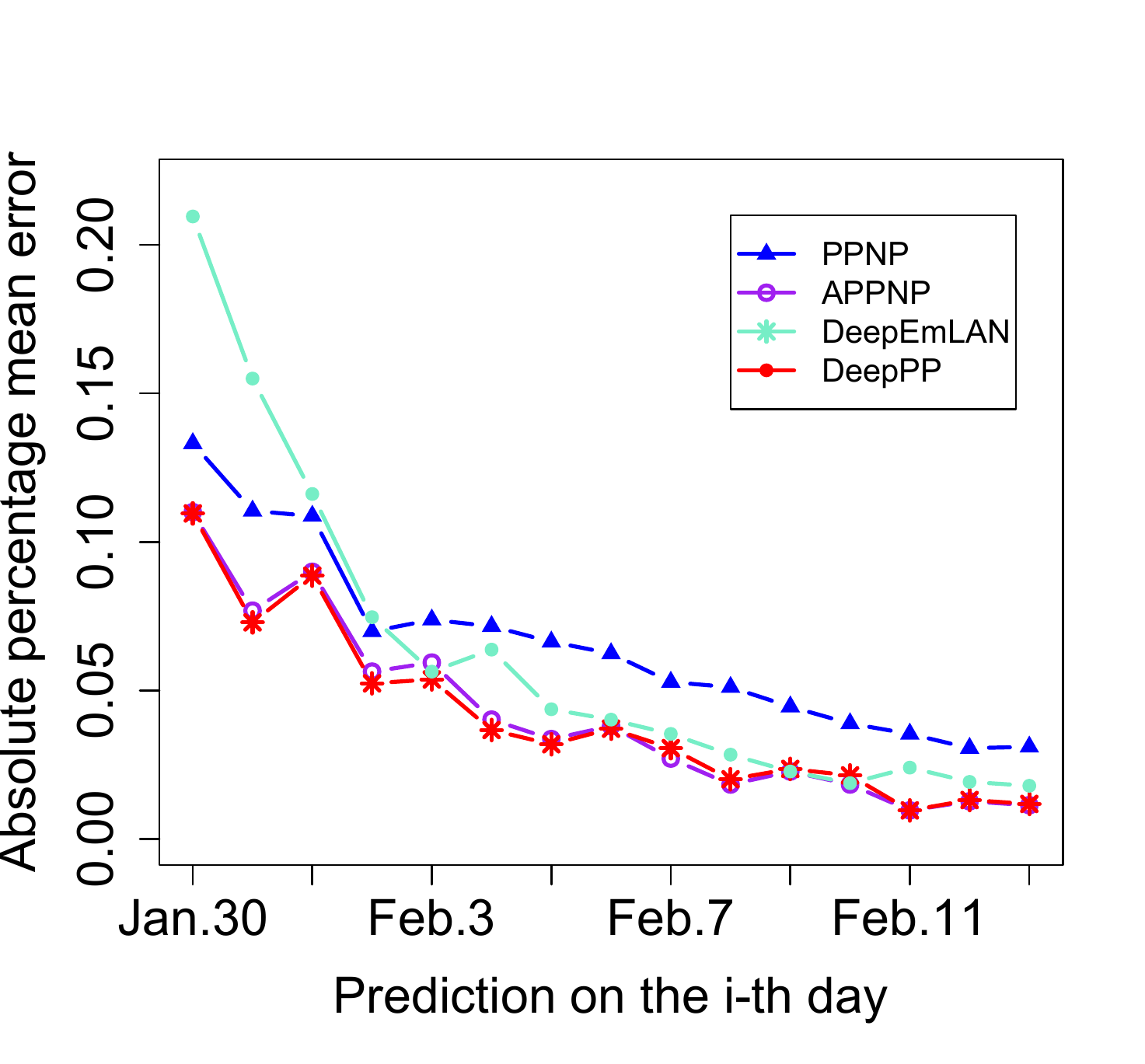}}
\subfigure[]{
\label{Fig10:sub:2}
\includegraphics[width=0.31\textwidth]{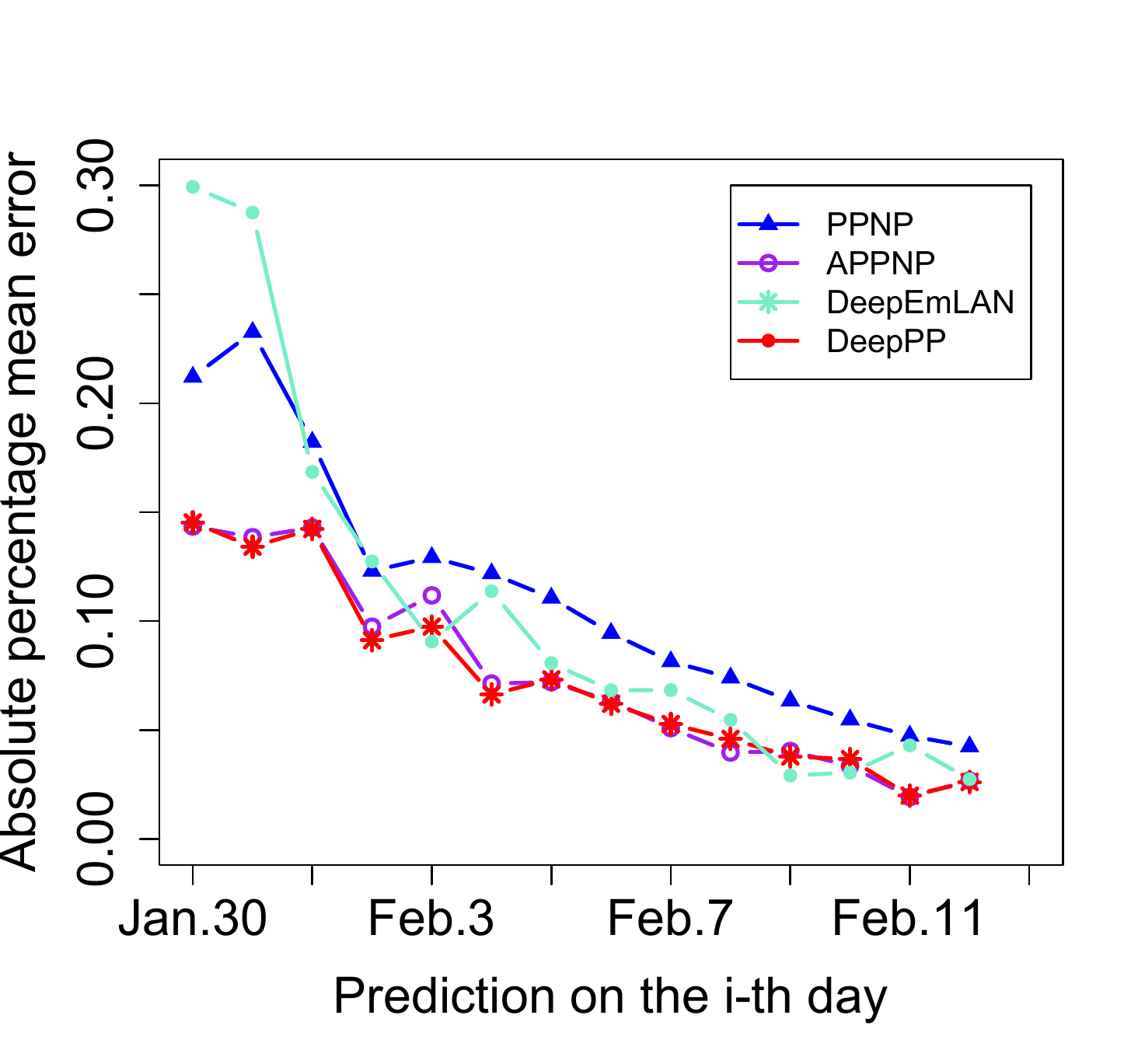}}
\subfigure[]{
\label{Fig10:sub:3}
\includegraphics[width=0.31\textwidth]{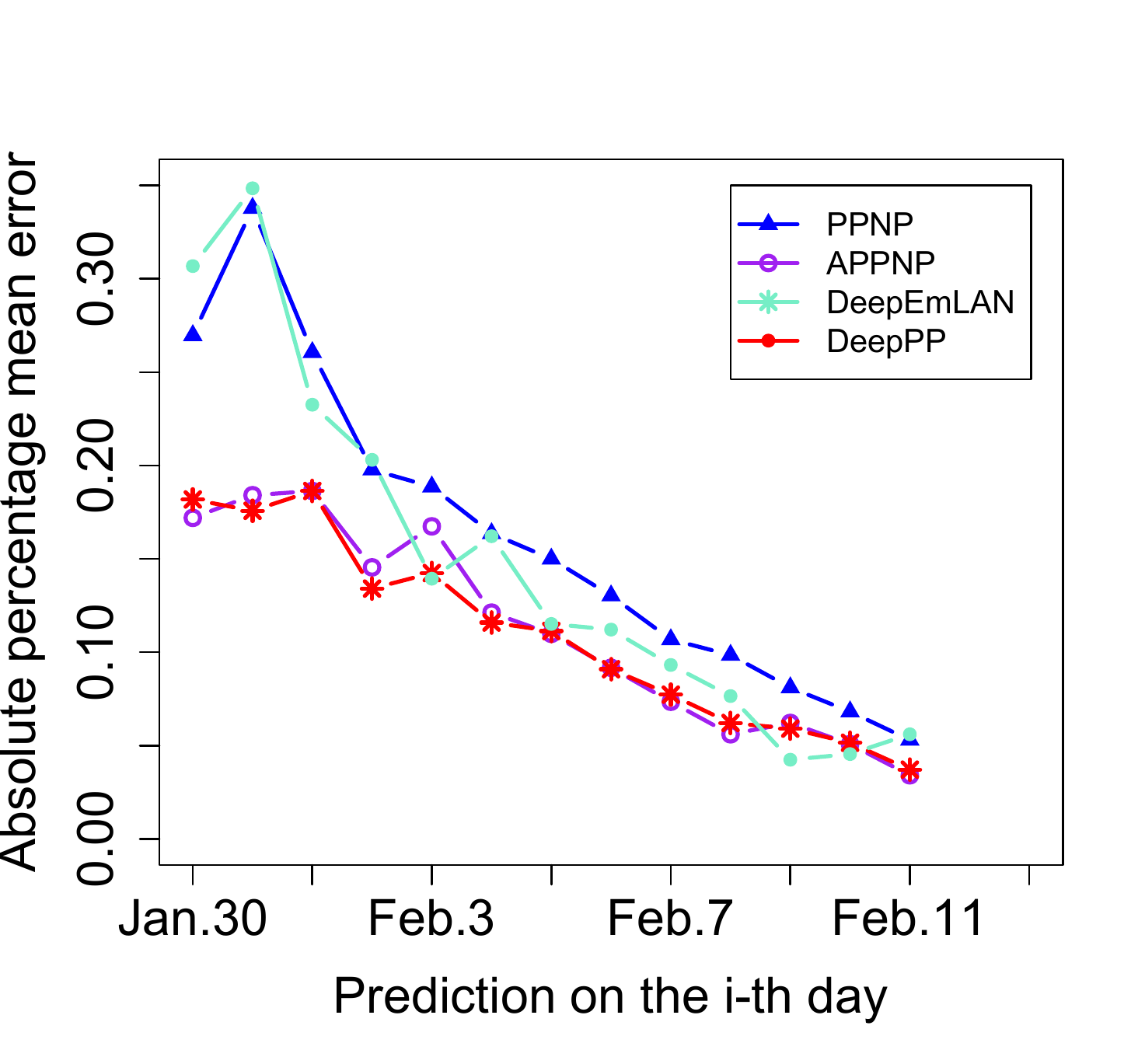}}
\subfigure[]{
\label{Fig10:sub:4}
\includegraphics[width=0.31\textwidth]{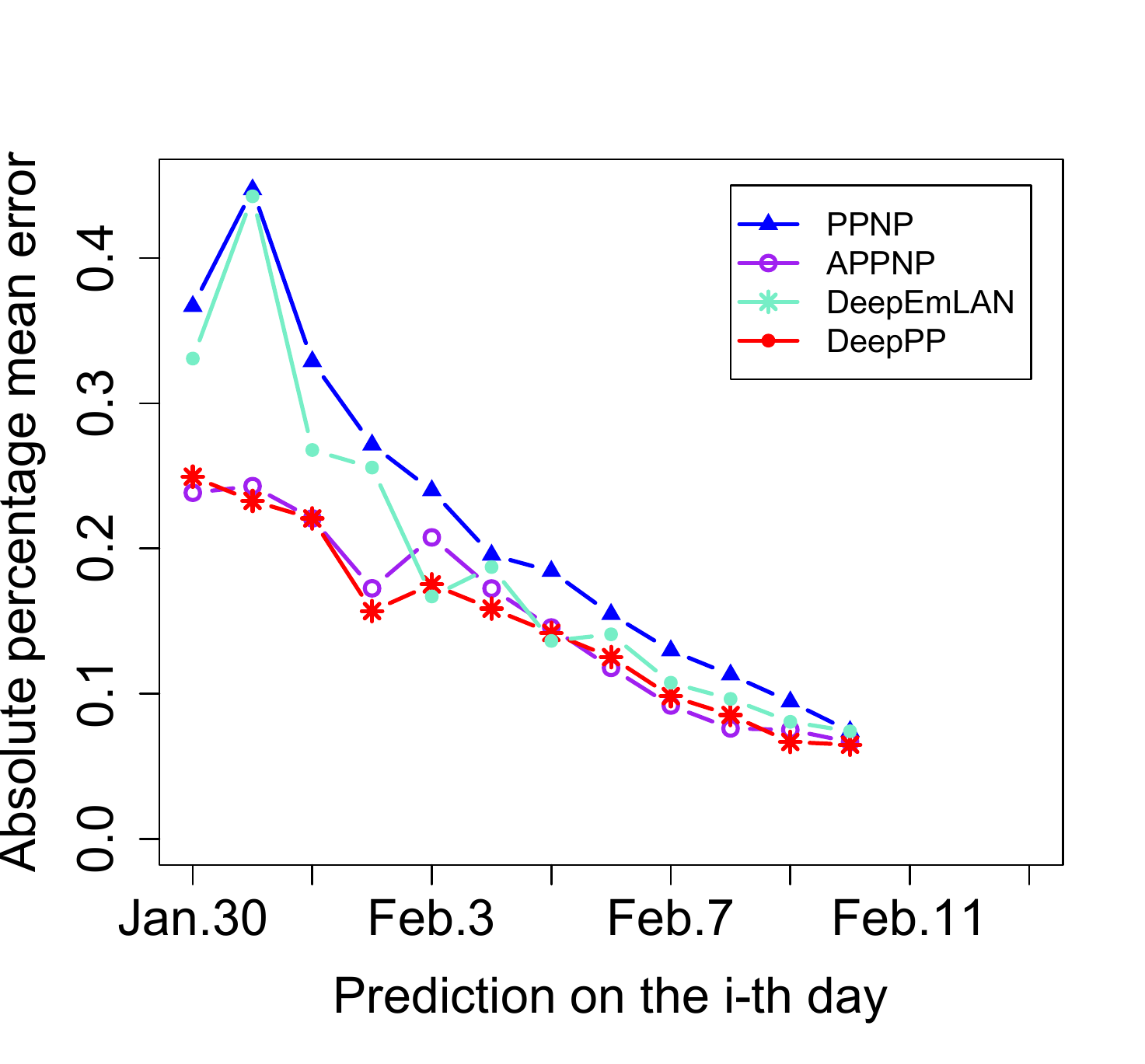}}
\subfigure[]{
\label{Fig10:sub:5}
\includegraphics[width=0.31\textwidth]{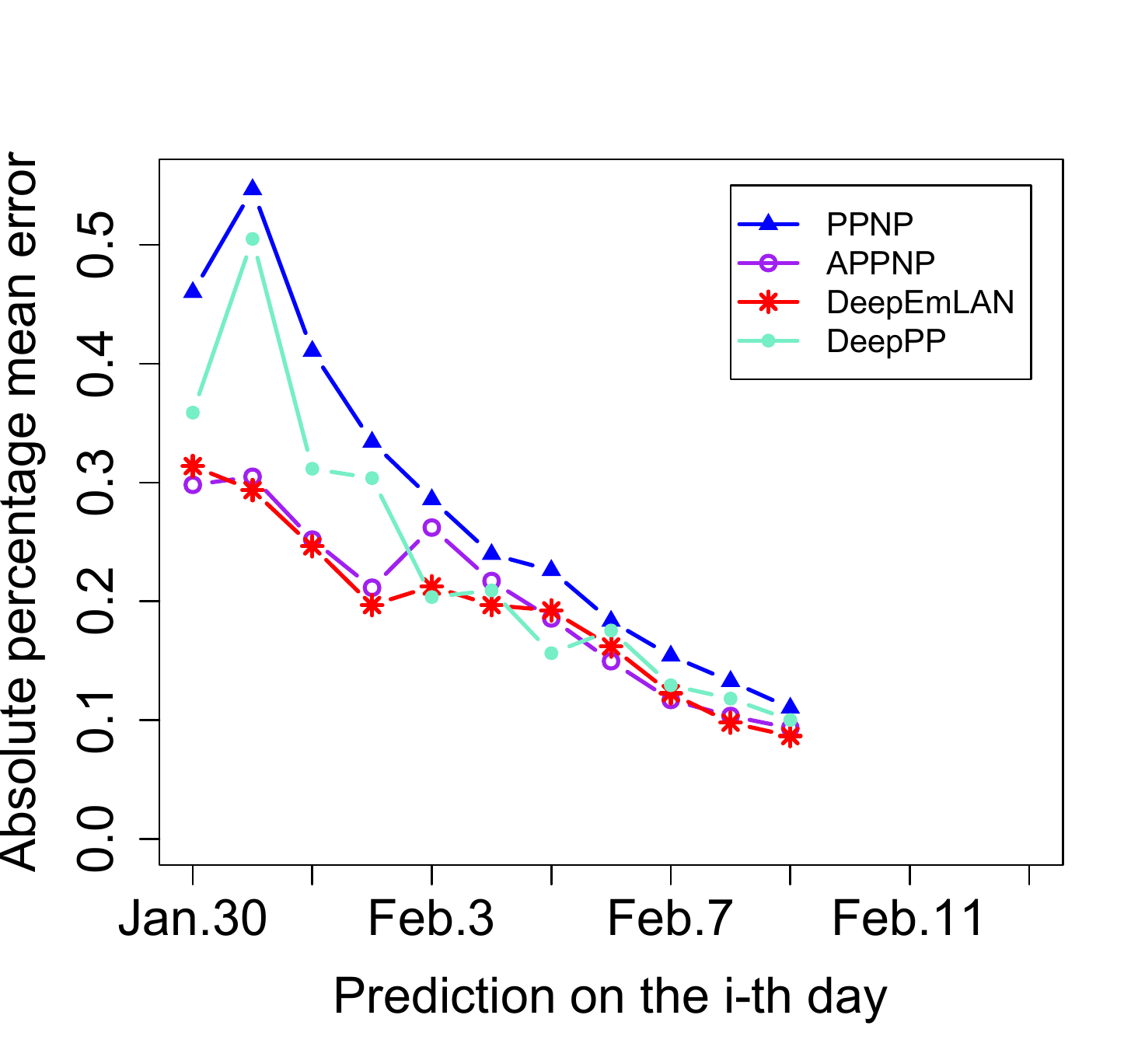}}
\subfigure[]{
\label{Fig10:sub:6}
\includegraphics[width=0.31\textwidth]{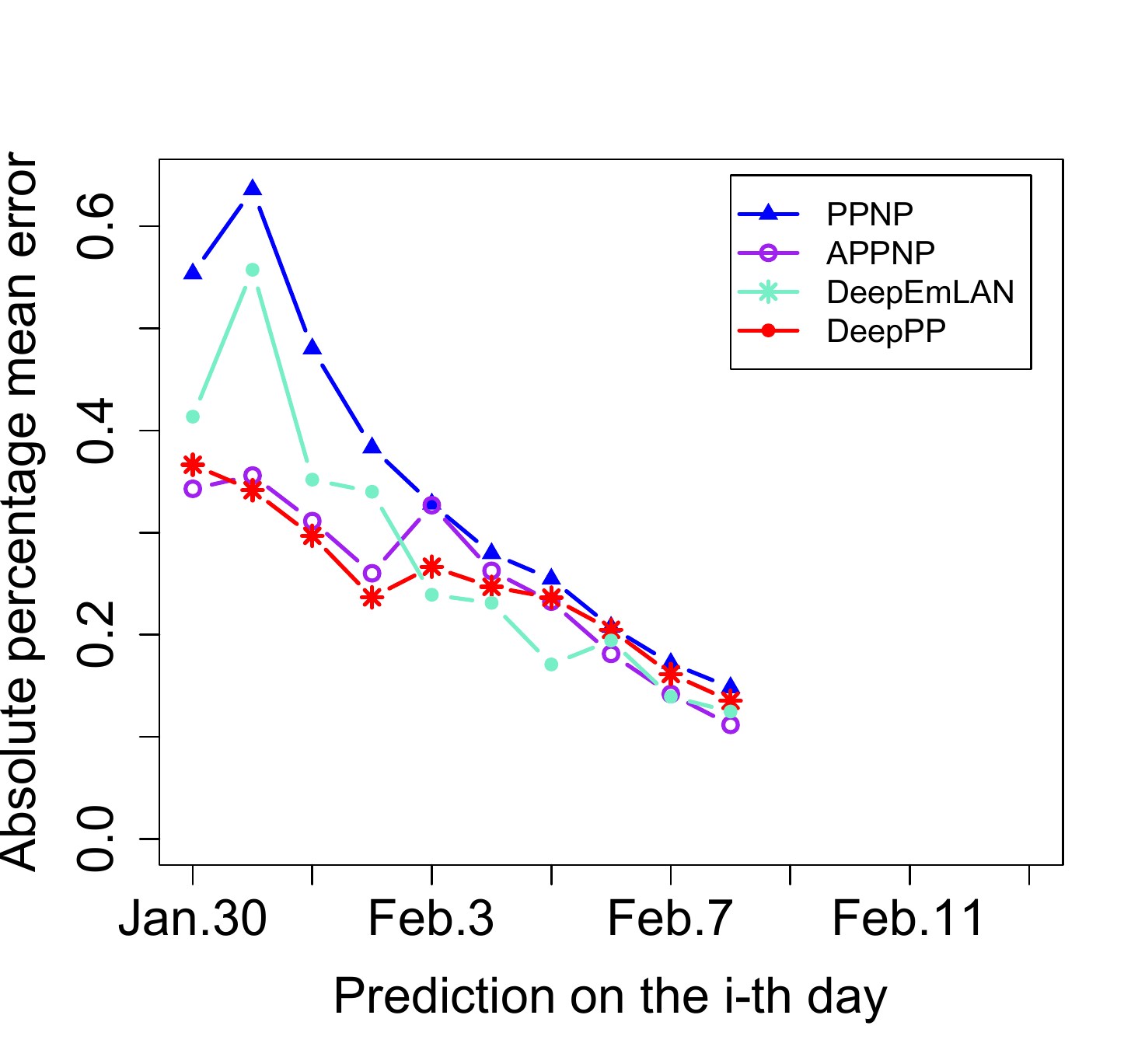}}
\caption{COVID-19 day-ahead predictions from Hubei China. Predictions are given over a 1-to-6-day interval (see subfigures (a) to (f)).}
\label{Fig:10}
\end{figure}

Figure \ref{Fig:10} shows the change in prediction accuracy of different prediction algorithms over time. The dates are displayed along the horizontal axis. From January 22, we used all the available information to make COVID-19 day-ahead prediction. For example, in Figure \ref{Fig10:sub:1}, the rightmost point displays the results from January 22 to February 13 to predict what will happen on February 14.

Over time, the absolute percentage mean error (APME) error has a tendency to decrease as the amount of data available increases. A rapid increase in infection cases was followed by a more gradual trend, displaying a sub exponential rise in daily infections. The APME error will decrease with sub-exponential growth since it is a measure of relative error. Also, as the prediction horizon extends, prediction accuracy decreases rapidly. As shown in Figures \ref{Fig10:sub:5} and \ref{Fig10:sub:6}, it is not possible to accurately predict the number of cases five or six days before and after February 1.

DeepEmLAN performed better, but it did not find an accurate prediction before and after January 31. The time series in the leftmost part of Figure \ref{Fig:10} is the shortest, so there was less data available to train DeepEmLAN. In these cases with short time series, the prediction accuracy of this pure machine learning algorithm was lower than the other methods.

\subsubsection{Holland}

\begin{figure}[!htbp]
\centering
\subfigure[]{
\label{Fig:sub:1}
\includegraphics[width=0.48\textwidth]{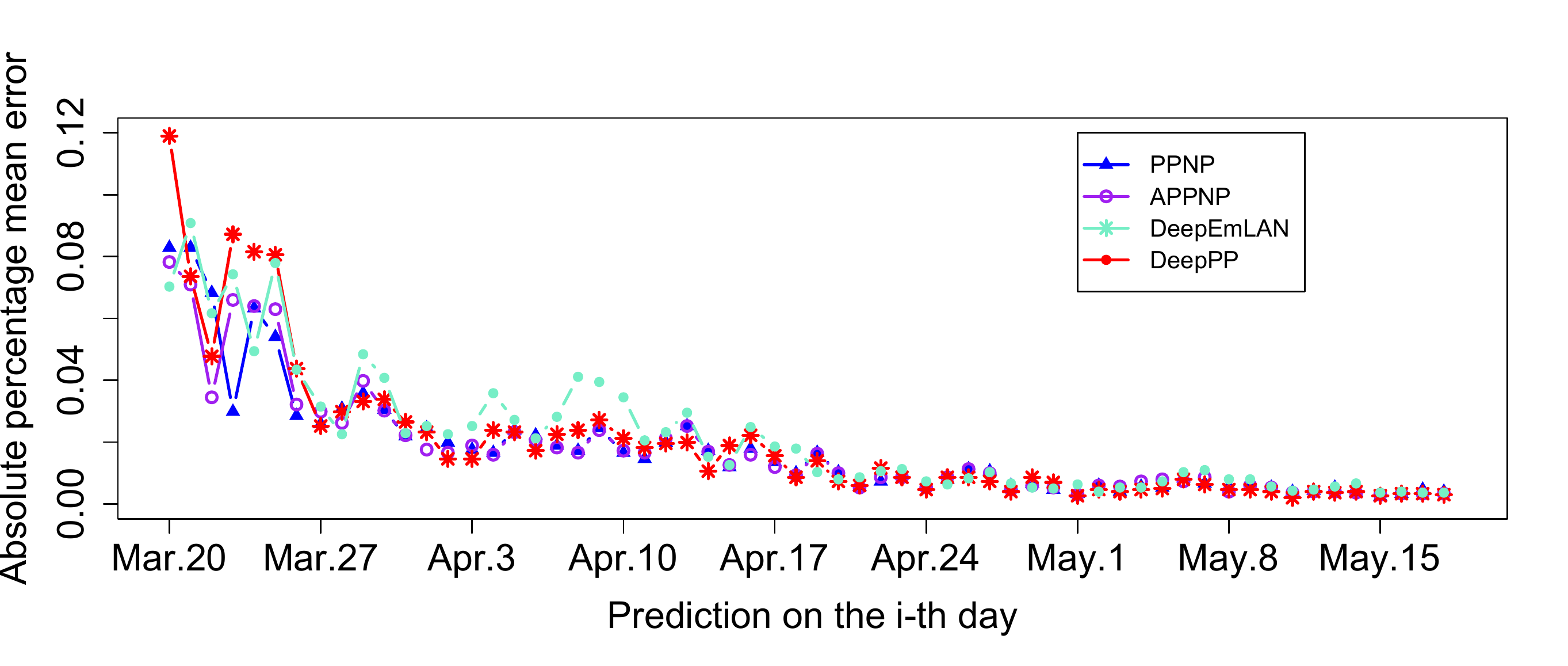}}
\subfigure[]{
\label{Fig:sub:2}
\includegraphics[width=0.48\textwidth]{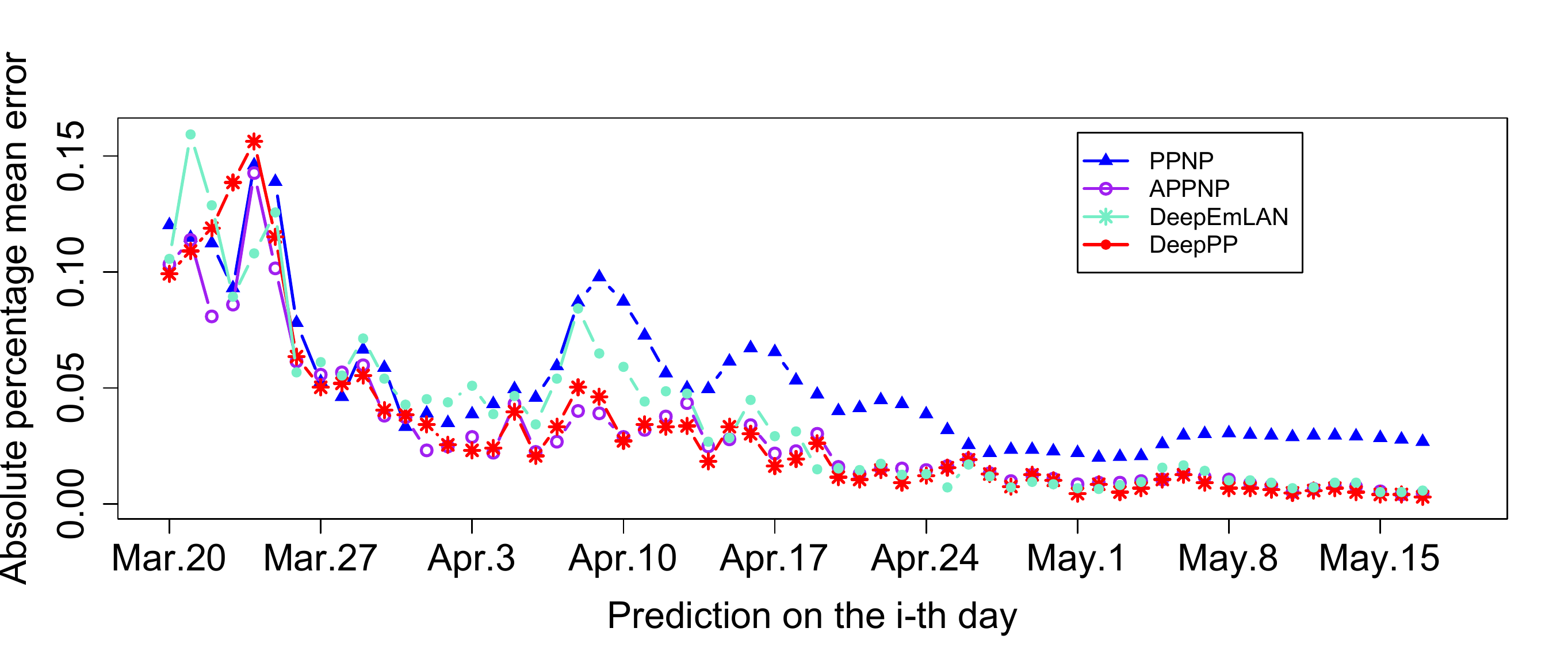}}
\subfigure[]{
\label{Fig:sub:3}
\includegraphics[width=0.48\textwidth]{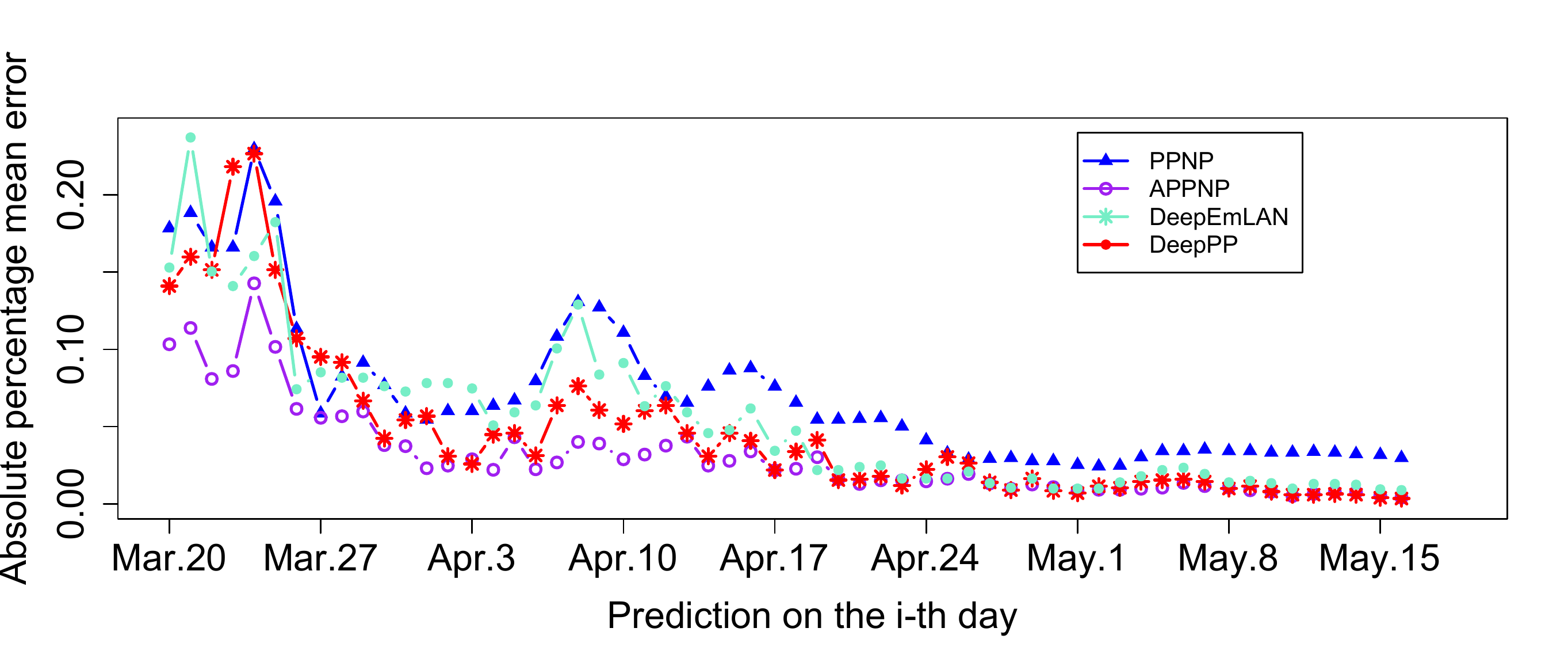}}
\subfigure[]{
\label{Fig:sub:4}
\includegraphics[width=0.48\textwidth]{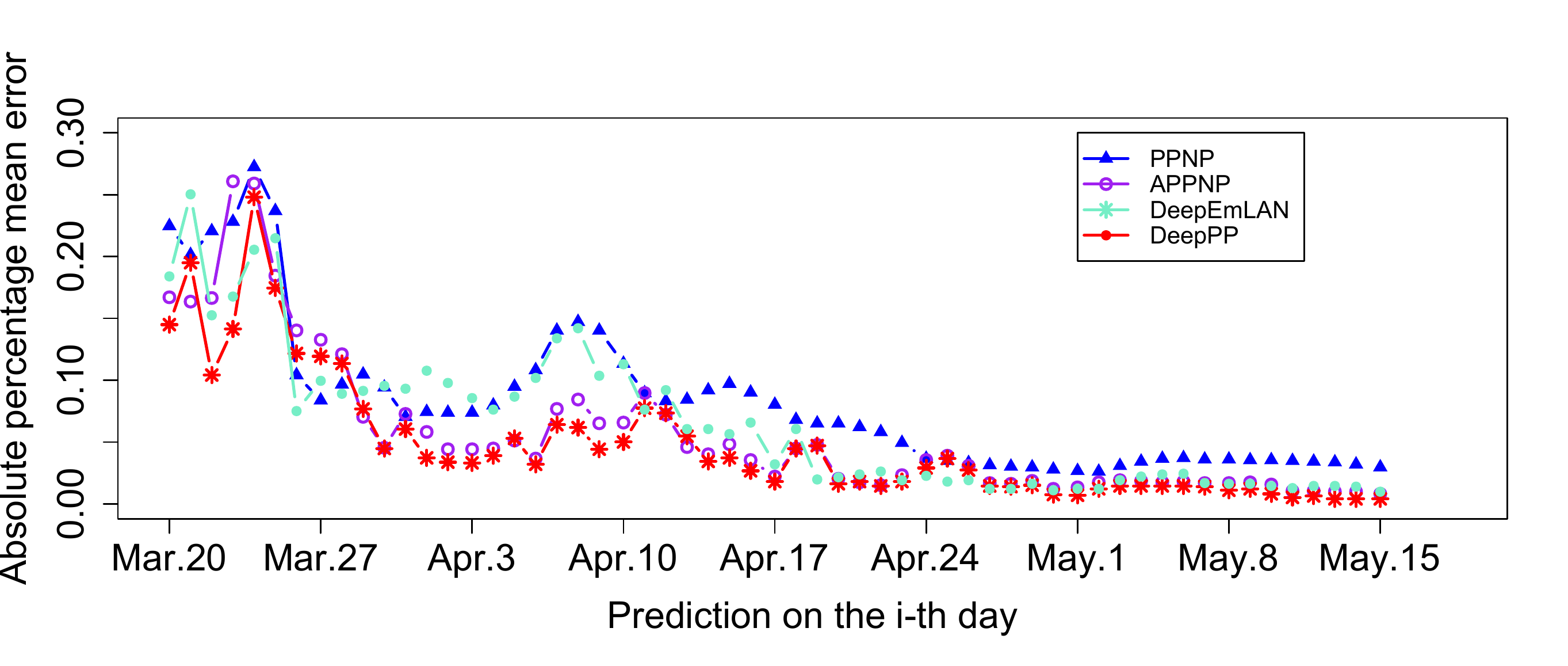}}
\subfigure[]{
\label{Fig:sub:5}
\includegraphics[width=0.48\textwidth]{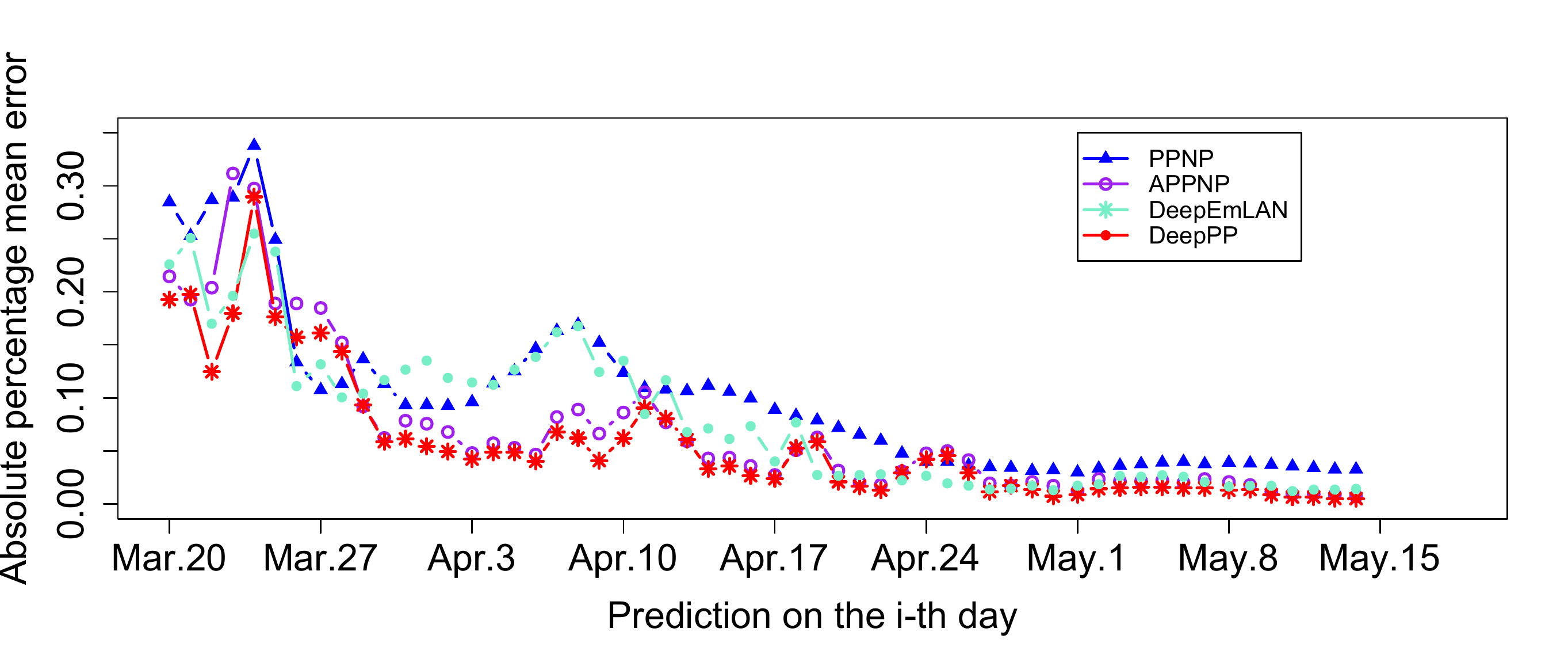}}
\subfigure[]{
\label{Fig:sub:6}
\includegraphics[width=0.48\textwidth]{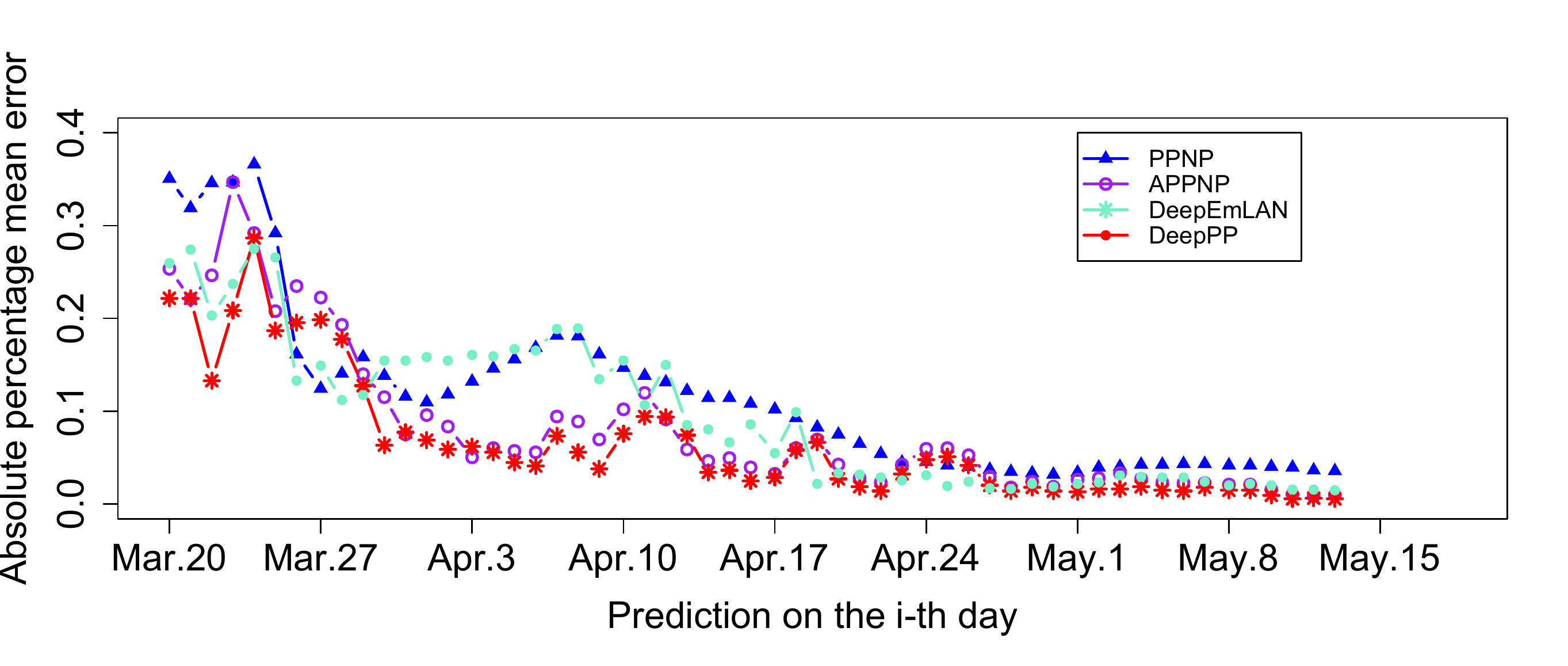}}
\caption{COVID-19 day-ahead predictions from Holland. Predictions are given over a 1-to-6-day interval (see subfigures (a) to (f)).}
\label{Fig:11}
\end{figure}

Prediction accuracy with the Holland dataset is shown in Figure \ref{Fig:11}. The COVID-19 situation in Holland was essentially the same as in Hubei prior to April 1, 2020. Here, DeepPP proved to be the best method but with large deviations in prediction accuracy. All compared algorithms have been roughly the same since April 1. Also, whether the network is initially static or dynamic appears to have little effect on prediction accuracy. The DeepPP algorithm is trained on more and more infection data over time. The DeepInf-GCN and DeepInf-GAT separation methods performed best in the whole cycle, while PPNP and DeepEmLAN performed worst.

The prediction accuracy of DeepPP and APPNP is comparable. One possible reason is that the transmission of COVID-19 was primarily inter provincial interaction. The COVID-19 spread mainly in the provinces after the end of March. We then compared the spread of COVID-19 across all seven algorithms. The errors are listed in Table \ref{table:4}, obtained by averaging all APME prediction errors over a 1-to-6-day prediction interval. Table \ref{table:4} clearly illustrates that the prediction error of DeepPP algorithm is smaller than the other algorithms, because DeepPP takes into account cities interact with each other. The prediction errors for DeepPP in each city are equal to that of DeepEmLAN. In conclusion, the network-based method offers better prediction accuracy.

\begin{table}[htbp]
	\centering
	\caption{Error comparison of different methods}
	\begin{tabular}{cccc}
		\toprule  % Â¶Â¥Â²Â¿ÃÃ
		Algorithm & Error in Hubei & Error in Holland & Bias\\
		\midrule  % ÃÃÂ²Â¿ÃÃ
		GNN* & 0.15 & 0.043 & under\\
        DeepInf-GCN & 0.14 & 0.042 & over\\
        DeepInf-GAT & 0.13 & 0.044 &  \\
        PPNP & 0.14 & 0.051 & over\\
        APPNP & 0.14 & 0.043 & \\
        DeepEmLAN & 0.16 & 0.057 & under\\
        DeepPP & 0.13 & 0.038 &  \\
		\bottomrule  % ÂµÃÂ²Â¿ÃÃ
	\end{tabular}
\label{table:4}
\end{table}

\subsubsection{Ablation experiments}

In the ablation experiments, we compared multiple variants of DeepPP: DeepPP-RNN, DeepPP-noPNF, DeepPP-noScoreAttention, DeepPP-noPointAdjust, using the DeepPP prototype as the control model. DeepPP-RNN replaces the TCN unit with an RNN. DeepPP-noPNF replaces the transformation process in the latent space with the standard Gaussian distribution $N(0,1)$. DeepPP-noScoreAttention omits the score attention mechanism and, similarly, DeepPP-noPointAdjust omits the point adjust method. For fairness, we also eliminated the peaks-over-threshold (POT) mechanism in our model and used $F1_{best}$ as the evaluation indicator.

\begin{figure}[!htbp]
\begin{center}
\includegraphics*[width=0.7\textwidth]{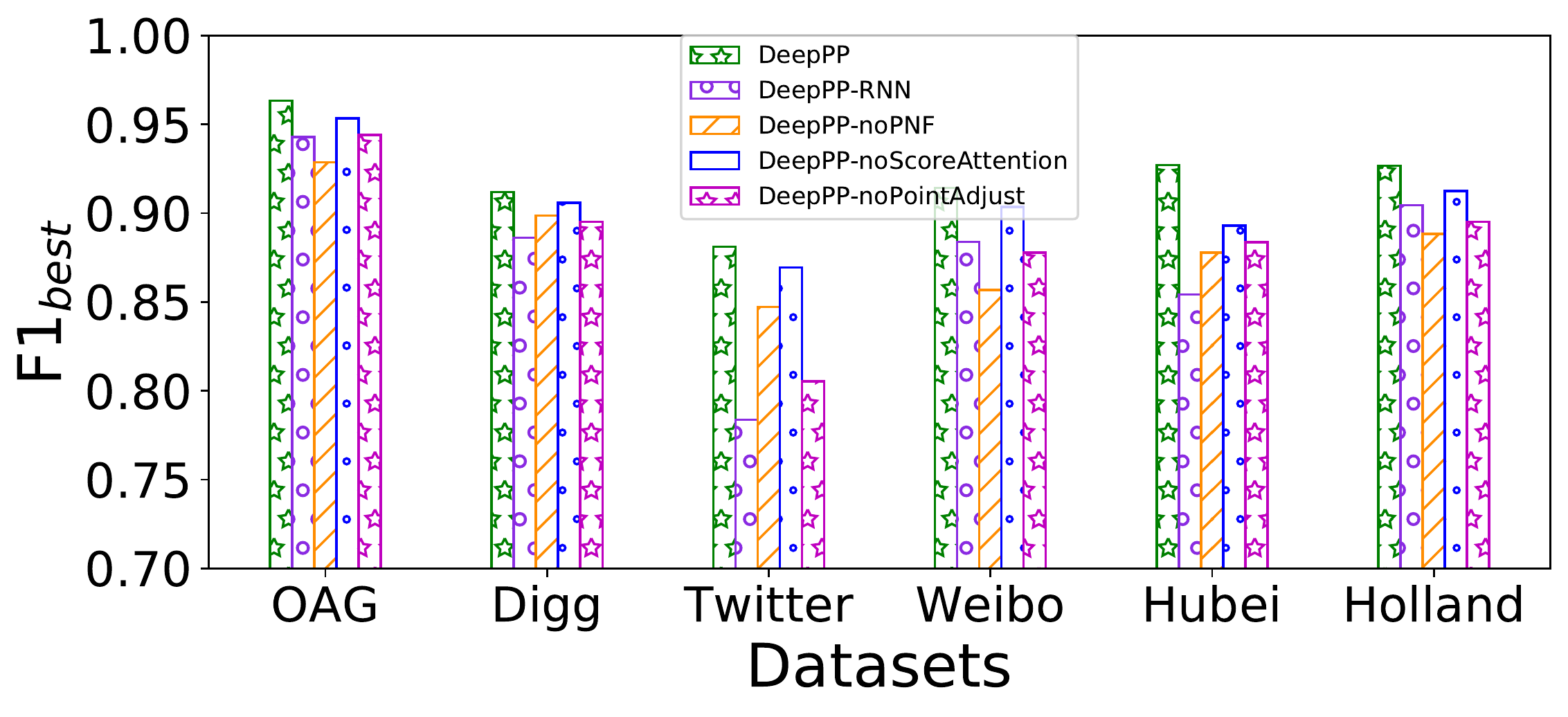}
\caption{A comparative analysis of ablation experiment in $F1_{best}$ on six datasets}
\label{fig:12}
\end{center}
\end{figure}

As can be seen from Figure \ref{fig:12}, DeepPP-RNN did not perform as well as DeepPP. This is because a simple RNN cannot capture the long-term dependencies in time series data. Moreover, DeepPP resulted in a much smaller model that took far less time to train compared to DeepPP-RNN. Hence, training DeepPP would be much easier. From the perspective of the impact of latent space, PNF can capture complex latent variable data patterns and help build and generate latent space variables. From DeepPP-noPNF and DeepPP-RNN in Figure \ref{fig:12}, it can be seen that the noPNF variant has less impact on the model than the RNN variant, while the RNN variant had a greater impact with the Digg dataset. This is because the scale of packet data in the Digg dataset is very small, so the RNN was able to capture the time dependencies.

In general, the impact of DeepPP-noScoreAttention and DeepPP-noPointAdjust on model organization was smaller than DeepPP-RNN and DeepPP-noPNF. After DeepPP-noScoreAttention eliminates the abnormal scoring mechanism, the gap between DeepPP-noScoreAttention and DeepPP is very small. This is because the score attention mechanism has abnormal data close to normal data. The detection results have an impact, but the impact is much smaller than with the overall model. DeepPP-noPointAdjust had the greatest impact with the Twitter dataset. This is because point adjustment can effectively detect continuous anomaly types. Compared to abnormally scattered datasets and massive data, the impact is much greater.

\section{Conclusion}
In this paper, we proposed a deep learning-based personalized propagation algorithm named DeepPP. By extending DeepInf, this algorithm integrates the transition probability of the page rank domain with a GCN. The method can adjust the size of a neighbor's influence. It also has greater flexibility. A variety of social networks (OAG, Digg, Twitter, and Weibo) as well as two COVID-19 datasets (Hubei and Holland) were studied in extensive experiments. As demonstrated by the results, DeepPP yielded better $F1-Measure$ than the current state-of-the-art methods. The proposed method performed better on the COVID-19 datasets in precision. The excellent performance of DeepPP on various datasets proves that it can be effectively applied to general real-world scenarios for predicting the social influence of COVID-19. However, although we used a transition probability to achieve greater flexibility in our modeling, we did not consider any user-specified constraints. Further exploring and incorporating these into the model is a worthy future research direction. Another exciting direction would be to use reinforcement learning to combine sampling and learning for modeling the social influence of COVID-19.

%\section*{Declaration of Interest Statement}
%The authors declare that they have no known competing financial interests or personal relationships that may appear to influence the work reported in this paper.

\section*{References}
\bibliography{elsarticle-template.bbl}

\begin{thebibliography}{10}
\expandafter\ifx\csname url\endcsname\relax
  \def\url#1{\texttt{#1}}\fi
\expandafter\ifx\csname urlprefix\endcsname\relax\def\urlprefix{URL }\fi
\expandafter\ifx\csname href\endcsname\relax
  \def\href#1#2{#2} \def\path#1{#1}\fi

\bibitem{LU2016143}
W.-X. Lu, C.~Zhou, J.~Wu, Big social network influence maximization via
  recursively estimating influence spread, Knowledge-Based Systems 113 (2016)
  143--154.

\bibitem{WANG2017154}
H.~Wang, J.~Wu, S.~Pan, P.~Zhang, L.~Chen, Towards large-scale social networks
  with online diffusion provenance detection, Computer Networks 114 (2017)
  154--166.

\bibitem{DAUD2020102716}
N.~N. Daud, S.~H. {Ab Hamid}, M.~Saadoon, F.~Sahran, N.~B. Anuar, Applications
  of link prediction in social networks: A review, Journal of Network and
  Computer Applications 166 (2020) 102716.

\bibitem{SINGH2021151}
A.~K. Singh, L.~Kailasam, Link prediction-based influence maximization in
  online social networks, Neurocomputing 453 (2021) 151--163.
\newblock \href
  {http://dx.doi.org/https://doi.org/10.1016/j.neucom.2021.04.084}
  {\path{doi:https://doi.org/10.1016/j.neucom.2021.04.084}}.

\bibitem{DU2022386}
H.~Du, Y.~Zhou, Nostradamus: A novel event propagation prediction approach with
  spatio-temporal characteristics in non-euclidean space, Neural Networks 145
  (2022) 386--394.

\bibitem{liu2018social}
C.-Y. Liu, C.~Zhou, J.~Wu, Y.~Hu, L.~Guo, Social recommendation with an
  essential preference space, in: Thirty-second AAAI conference on artificial
  intelligence, 2018.

\bibitem{Gao_CIKM16}
L.~Gao, J.~Wu, Z.~Qiao, C.~Zhou, H.~Yang, Y.~Hu, Collaborative social group
  influence for event recommendation, in: Proceedings of the 25th ACM
  International on Conference on Information and Knowledge Management, CIKM
  '16, 2016, p. 1941–1944.

\bibitem{RAJ202236}
C.~Raj, P.~Meel, Arcnn framework for multimodal infodemic detection, Neural
  Networks 146 (2022) 36--68.

\bibitem{YANG2021107640}
Z.~Yang, Q.~Li, H.~Xie, Q.~Wang, W.~Liu, Learning representation from multiple
  media domains for enhanced event discovery, Pattern Recognition 110 (2021)
  107640.

\bibitem{ijcai2020-693}
F.~Liu, S.~Xue, J.~Wu, C.~Zhou, W.~Hu, C.~Paris, S.~Nepal, J.~Yang, P.~S. Yu,
  Deep learning for community detection: Progress, challenges and
  opportunities, in: Proceedings of the Twenty-Ninth International Joint
  Conference on Artificial Intelligence, {IJCAI}, 2020, pp. 4981--4987.

\bibitem{ma2021comprehensive}
X.~Ma, J.~Wu, S.~Xue, J.~Yang, C.~Zhou, Q.~Z. Sheng, H.~Xiong, L.~Akoglu, A
  comprehensive survey on graph anomaly detection with deep learning, IEEE
  Transactions on Knowledge and Data Engineering.

\bibitem{matsubara2012rise}
Y.~Matsubara, Y.~Sakurai, B.~A. Prakash, L.~Li, C.~Faloutsos, Rise and fall
  patterns of information diffusion: model and implications, in: Proceedings of
  the 18th ACM SIGKDD international conference on Knowledge discovery and data
  mining, 2012, pp. 6--14.

\bibitem{li2017deepcas}
C.~Li, J.~Ma, X.~Guo, Q.~Mei, Deepcas: An end-to-end predictor of information
  cascades, in: Proceedings of the 26th international conference on World Wide
  Web, 2017, pp. 577--586.

\bibitem{qiu2018deepinf}
J.~Qiu, J.~Tang, H.~Ma, Y.~Dong, K.~Wang, J.~Tang, Deepinf: Social influence
  prediction with deep learning, in: Proceedings of the 24th ACM SIGKDD
  International Conference on Knowledge Discovery \& Data Mining, 2018, pp.
  2110--2119.

\bibitem{klicpera2018personalized}
J.~Klicpera, A.~Bojchevski, S.~G{\"u}nnemann, Personalized embedding
  propagation: Combining neural networks on graphs with personalized pagerank,
  CoRR, abs/1810.05997.

\bibitem{4700287}
F.~Scarselli, M.~Gori, A.~C. Tsoi, M.~Hagenbuchner, G.~Monfardini, The graph
  neural network model, IEEE Transactions on Neural Networks 20~(1) (2009)
  61--80.

\bibitem{ZHAO2021382}
Z.~Zhao, H.~Zhou, C.~Li, J.~Tang, Q.~Zeng, Deepemlan: Deep embedding learning
  for attributed networks, Information Sciences 543 (2021) 382--397.

\bibitem{li2018social}
K.~Li, L.~Zhang, H.~Huang, Social influence analysis: Models, methods, and
  evaluation, Engineering 4~(1) (2018) 40--46, cybersecurity.

\bibitem{Koskinen}
J.~Koskinen, G.~Daraganova, Bayesian analysis of social influence (2020).

\bibitem{WU20151487}
J.~Wu, S.~Pan, X.~Zhu, Z.~Cai, P.~Zhang, C.~Zhang, Self-adaptive attribute
  weighting for naive bayes classification, Expert Systems with Applications
  42~(3) (2015) 1487--1502.

\bibitem{wu2011attribute}
J.~Wu, Z.~Cai, Attribute weighting via differential evolution algorithm for
  attribute weighted naive bayes (wnb), Journal of Computational Information
  Systems 7~(5) (2011) 1672--1679.

\bibitem{liao2018attributed}
L.~Liao, X.~He, H.~Zhang, T.-S. Chua, Attributed social network embedding, IEEE
  Transactions on Knowledge and Data Engineering 30~(12) (2018) 2257--2270.

\bibitem{wang2019heterogeneous}
X.~Wang, H.~Ji, C.~Shi, B.~Wang, Y.~Ye, P.~Cui, P.~S. Yu, Heterogeneous graph
  attention network, in: WWW, 2019, pp. 2022--2032.

\bibitem{wu2019simplifying}
F.~Wu, A.~Souza, T.~Zhang, C.~Fifty, T.~Yu, K.~Weinberger, Simplifying graph
  convolutional networks, in: International conference on machine learning,
  PMLR, 2019, pp. 6861--6871.

\bibitem{zhang2018salient}
Q.~Zhang, J.~Wu, P.~Zhang, G.~Long, C.~Zhang, Salient subsequence learning for
  time series clustering, IEEE transactions on pattern analysis and machine
  intelligence 41~(9) (2018) 2193--2207.

\bibitem{wu2019dual}
Q.~Wu, H.~Zhang, X.~Gao, P.~He, P.~Weng, H.~Gao, G.~Chen, Dual graph attention
  networks for deep latent representation of multifaceted social effects in
  recommender systems, in: The World Wide Web Conference, 2019, pp. 2091--2102.

\bibitem{cao2017deephawkes}
Q.~Cao, H.~Shen, K.~Cen, W.~Ouyang, X.~Cheng, Deephawkes: Bridging the gap
  between prediction and understanding of information cascades, in: Proceedings
  of the 2017 ACM on Conference on Information and Knowledge Management, 2017,
  pp. 1149--1158.

\bibitem{menon2018proper}
A.~Menon, Y.~Lee, Proper loss functions for nonlinear hawkes processes, in:
  Proceedings of the AAAI Conference on Artificial Intelligence, Vol.~32, 2018.

\bibitem{gou2018learning}
C.~Gou, H.~Shen, P.~Du, D.~Wu, Y.~Liu, X.~Cheng, Learning sequential features
  for cascade outbreak prediction, Knowledge and Information Systems 57~(3)
  (2018) 721--739.

\bibitem{kefato2018cas2vec}
Z.~T. Kefato, N.~Sheikh, L.~Bahri, A.~Soliman, A.~Montresor, S.~Girdzijauskas,
  Cas2vec: Network-agnostic cascade prediction in online social networks, in:
  2018 Fifth International Conference on Social Networks Analysis, Management
  and Security (SNAMS), IEEE, 2018, pp. 72--79.

\bibitem{perozzi2014deepwalk}
B.~Perozzi, R.~Al-Rfou, S.~Skiena, Deepwalk: Online learning of social
  representations, in: Proceedings of the 20th ACM SIGKDD international
  conference on Knowledge discovery and data mining, 2014, pp. 701--710.

\bibitem{wang2017community}
X.~Wang, P.~Cui, J.~Wang, J.~Pei, W.~Zhu, S.~Yang, Community preserving network
  embedding, in: AAAI, 2017, pp. 203--–209.

\bibitem{grover2016node2vec}
A.~Grover, J.~Leskovec, node2vec: Scalable feature learning for networks, in:
  Proceedings of the 22nd ACM SIGKDD international conference on Knowledge
  discovery and data mining, 2016, pp. 855--864.

\bibitem{dong2017metapath2vec}
Y.~Dong, N.~V. Chawla, A.~Swami, metapath2vec: Scalable representation learning
  for heterogeneous networks, in: Proceedings of the 23rd ACM SIGKDD
  international conference on knowledge discovery and data mining, 2017, pp.
  135--144.

\bibitem{liu2019general}
X.~Liu, T.~Murata, K.-S. Kim, C.~Kotarasu, C.~Zhuang, A general view for
  network embedding as matrix factorization, in: Proceedings of the Twelfth ACM
  International Conference on Web Search and Data Mining, 2019, pp. 375--383.

\bibitem{siglidis2020grakel}
G.~Siglidis, G.~Nikolentzos, S.~Limnios, C.~Giatsidis, K.~Skianis,
  M.~Vazirgiannis, Grakel: A graph kernel library in python., Journal of
  Machine Learning Research 21~(54) (2020) 1--5.

\bibitem{niepert2016learning}
M.~Niepert, M.~Ahmed, K.~Kutzkov, Learning convolutional neural networks for
  graphs, in: International conference on machine learning, PMLR, 2016, pp.
  2014--2023.

\bibitem{Hamilton_NIPS}
W.~L. Hamilton, R.~Ying, J.~Leskovec, Inductive representation learning on
  large graphs, in: Proceedings of the 31st International Conference on Neural
  Information Processing Systems, NIPS'17, 2017, p. 1025–1035.

\bibitem{ksantini2007weighted}
R.~Ksantini, D.~Ziou, B.~Colin, F.~Dubeau, Weighted pseudometric discriminatory
  power improvement using a bayesian logistic regression model based on a
  variational method, IEEE transactions on pattern analysis and machine
  intelligence 30~(2) (2007) 253--266.

\bibitem{song2019session}
W.~Song, Z.~Xiao, Y.~Wang, L.~Charlin, M.~Zhang, J.~Tang, Session-based social
  recommendation via dynamic graph attention networks, in: Proceedings of the
  Twelfth ACM International Conference on Web Search and Data Mining, 2019, pp.
  555--563.

\bibitem{martin2021attention}
A.~Martin, S.~I. Becker, A.~J. Pegna, Attention is prioritised for proximate
  and approaching fearful faces, Cortex 134 (2021) 52--64.

\bibitem{batagelj2001cores}
V.~Batagelj, M.~Zavernik, Cores decomposition of networks, Recent Trends in
  Graph Theory, Algebraic Combinatorics, and Graph Algorithms (2001) 24--27.

\bibitem{schank2005approximating}
T.~Schank, D.~Wagner, Approximating clustering coefficient and transitivity.,
  Journal of Graph Algorithms and Applications 9~(2) (2005) 265--275.

\bibitem{zhang2015influenced}
J.~Zhang, J.~Tang, J.~Li, Y.~Liu, C.~Xing, Who influenced you? predicting
  retweet via social influence locality, ACM Transactions on Knowledge
  Discovery from Data (TKDD) 9~(3) (2015) 1--26.

\bibitem{LI2020107206}
H.~Li, X.~Liu, T.~Li, R.~Gan, A novel density-based clustering algorithm using
  nearest neighbor graph, Pattern Recognition 102 (2020) 107206.

\bibitem{ugander2012structural}
J.~Ugander, L.~Backstrom, C.~Marlow, J.~Kleinberg, Structural diversity in
  social contagion, Proceedings of the National Academy of Sciences 109~(16)
  (2012) 5962--5966.

\bibitem{10.1145/2835776.2835849}
J.~Tang, Aminer: Toward understanding big scholar data, in: Proceedings of the
  Ninth ACM International Conference on Web Search and Data Mining, WSDM '16,
  Association for Computing Machinery, New York, NY, USA, 2016, p. 467.

\bibitem{zhuang2018dual}
C.~Zhuang, Q.~Ma, Dual graph convolutional networks for graph-based
  semi-supervised classification, in: Proceedings of the 2018 World Wide Web
  Conference, 2018, pp. 499--508.

\bibitem{2020Wuhan}
Z.~Zhang, W.~Yao, Y.~Wang, C.~Long, X.~Fu, Wuhan and hubei covid-19 mortality
  analysis reveals the critical role of timely supply of medical resources,
  Journal of Infection 81~(1) (2020) 147--178.

\bibitem{2021Maternal}
J.~Guo, P.~D. Carli, P.~Lodder, M.~J. Bakermans-Kranenburg, M.~M.~E. Riem,
  Maternal mental health during the covid-19 lockdown in china, italy, and the
  netherlands: A cross-validation study, Psychological Medicine (2021) 1--44.

\bibitem{Wu_IJCAT}
J.~Wu, Z.-h. Cai, S.~Ao, Hybrid dynamic k-nearest-neighbour and distance and
  attribute weighted method for classification, Int. J. Comput. Appl. Technol.
  43~(4) (2012) 378–384.

\end{thebibliography}

\end{document}